\documentclass[10pt]{article}
\pdfoutput=1
\usepackage{amssymb,amsmath,mathrsfs,enumerate}
\usepackage{graphicx,rotate,multicol}
\usepackage{wrapfig}
\usepackage[colorlinks=true,
            linkcolor=red,
            urlcolor=blue,
            citecolor=blue]{hyperref}
\usepackage{color}
\usepackage{soul}
\usepackage{cite}
\usepackage{verbatim}
\long\def\rpl#1!!#2!!{\textcolor{red}{#1} \textcolor{blue}{#2}}

\usepackage{cleveref}
\usepackage[makeroom]{cancel}
\usepackage[normalem]{ulem}
\usepackage{amsmath,amssymb}
\usepackage{slashed}
\usepackage{xcolor}
\usepackage{graphicx}
\usepackage{cite}
\usepackage{pdflscape}
\usepackage{multirow}
\usepackage[thinlines]{easytable}
\usepackage{url}
\usepackage[utf8]{inputenc}
\usepackage{longtable}
\usepackage{booktabs}
\usepackage{graphicx,floatflt,amssymb,rotate, cancel}
\usepackage{mathrsfs}
\usepackage{amssymb}
\usepackage{amsmath}
\usepackage{subcaption}
\usepackage{float}
\usepackage{pstricks}
\usepackage[toc,page]{appendix}
\usepackage[mathscr]{euscript}
\usepackage{bbold}
\usepackage{color}

\def \order(#1){{\cal O} \left(#1 \right)}

\evensidemargin=\oddsidemargin
\allowdisplaybreaks

\begin{document}

\begin{flushright}
HRI-RECAPP-2020-009
\end{flushright}

\begin{center}
	{\Large \bf Revisiting Generalized Two Higgs Doublet Model in the Light of Muon Anomaly and Lepton Flavor Violating Decays at HL-LHC } \\
	\vspace*{1cm} {\sf Nivedita
          Ghosh$^{a,}$\footnote{niveditaghosh@hri.res.in},~Jayita
          Lahiri$^{a,}$\footnote{jayitalahiri@hri.res.in}} \\
	\vspace{10pt} {\small \em 
          $^a$Regional Centre for Accelerator-based Particle Physics,
Harish-Chandra Research Institute, HBNI,
Chhatnag Road, Jhunsi, Allahabad - 211 019, India}
	
	\normalsize
\end{center}

%
%
%
%
%
%
%
%


\begin{abstract}
One of the main motivations to look beyond the SM is the discrepancy between the theoretical prediction and observation of anomalous 
magnetic moment of muon. Alleviating this tension between theory and experiment and satisfying the bounds from lepton flavor violation data
simultaneously is a challenging task. In this paper, we consider generalised Two Higgs Doublet Model, with a Yukawa structure as a perturbation of Type X Two Higgs Doublet Model. In view of this model, we explore muon anomaly and lepton flavor violation along with constraints coming from B-physics, theoretical constraints, electroweak observables and collider data which can restrict the model parameter space significantly. We find that within the framework of this model it is possible to obtain regions allowed by all constraints, that can provide an explanation for the observed muon anomaly and at the same time predicts interesting signatures of lepton flavor violation.
 Furthermore, we consider the flavor violating decay of low-mass CP-odd scalar to probe the allowed parameter space at future runs of the LHC. With simple cut-based analysis 
we show that part of that parameter space can be probed with significance $> 5 \sigma$. We also provide Artificial Neural Network analysis which definitely improves our cut-based results significantly.

\end{abstract}

\bigskip

\section{Introduction}
A 125 GeV scalar, with a striking resemblance to the Higgs boson proposed in the Standard Model(SM) 
has been observed at the Large Hadron Collider(LHC)
~\cite{Aad:2012tfa,Chatrchyan:2012xdj} on July 4, 2012. Experimental evidence of its increasing biasness towards the SM puts stringent limits 
on New Physics(NP) scenarios. However, there are still many unanswered questions which indicate that SM is not a complete theory and it is imperative to go beyond it. Anomalous magnetic moment of muon is one such crucial observation which calls for new physics. There is a long-standing discrepancy between the SM prediction of magnetic moment of muon and its observed value in the experiment~\cite{Blum:2013xva}.
Ongoing E989 experiment at Fermilab~\cite{Grange:2015fou} and future E34 experiment at J-PARC~\cite{Iinuma:2011zz} are expected to shed new light on this tension between the theory and data. 

On the other hand, though lepton flavor violation(LFV) has been observed in neutrino oscillation experiments~\cite{Fukuda:1998mi,Ahmad:2002jz}, LFV has not yet been observed in the charged lepton sector. Various low energy experiments~\cite{Bellgardt:1988qe,Wintz:1998rp,Kuno:1999jp,Aubert:2009ag,TheMEG:2016wtm,Lindner:2016bgg,Bartolotta:2017mff,Beringer:1900zz} have put strong bounds on branching ratios of flavor changing neutral current(FCNC) decays and correspondingly on the associated couplings.

The two phenomena described above, namely, the muon anomaly and lepton flavor violation are intricately connected with each other. Experimental observation of muon anomaly and non-observation of lepton flavor violation will definitely create a tension in terms of the allowed parameter space for various candidate models which satisfy these two results individually. The models that can accommodate muon anomaly in general face severe LFV constraints~\cite{Lindner:2016bgg,Li:2019xmi}. In this work our goal is to satisfy both of these observations simultaneously and also, to look for signatures of lepton flavor violation in the collider experiments which is an independent search strategy altogether~\cite{Banerjee:2016foh,Primulando:2016eod,Primulando:2019ydt,Jana:2020pxx}. If in the future 
LFV is observed in low energy experiments, a simultaneous observation of LFV process in the collider will be a tell-tale signature of it. On the other hand, non-observation of LFV process at the collider will help us to constrain the model parameter space. 

In this work we consider a minimal scalar extension of the SM, i.e., generalised 
Two Higgs Doublet Model(2HDM)~\cite{Mahmoudi:2009zx,DiazCruz:2010yq,Bai:2012ex}, with a Yukawa structure as a perturbation of Type X 2HDM. The presence of non-standard light scalars in 2HDM allows one to satisfy muon anomaly which can be found in literature~\cite{Arhrib:2011wc,Liu:2015oaa,Chakrabarty:2018qtt,Iguro:2019sly,Chun:2019oix,Frank:2020smf,Chun:2020uzw,Broggio:2014mna,Cao:2009as,Wang:2014sda,Ilisie:2015tra,Abe:2015oca,Han:2015yys,Chun:2016hzs,Cherchiglia:2016eui,Cherchiglia:2017uwv,Wang:2018hnw,Chun:2015hsa}. The theory and phenomenology of FCNC in the Yukawa sector of 2HDM has also been studied in detail in the literature~\cite{Atwood:1996vj,Diaz:2000cm,Xiao:2003ya,Omura:2015xcg,Davidson:2016utf,Arhrib:2017yby}. However, a study of 2HDM 
in the light of both muon anomaly and LFV is scarce in the literature~\cite{DiazRodolfo:2000yy,Omura:2015nja}.

Apart from finding a suitable region of parameter space, where both muon anomaly and LFV constraints are satisfied in two loop along with theoretical constraints coming from perturbativity, unitarity, vacuum stability, oblique parameter constraints and constraints coming from B-physics and collider experiments, we look for a flavor violating decay of the 
CP-odd scalar in the $\ell^{+}\ell'^{-} + \slashed{E_T}$ final state, where $\ell, \ell'= e,\mu$. This channel comes from the flavor violating decay of the CP-odd Higgs, $A \to \ell \tau_{\ell'}$, where $\tau_{\ell'}$ implies $\tau$ decaying leptonically. 
With simple cut-based analysis we show that a selected region of parameter space can be probed in the future high luminosity collider HL-LHC. We also perform an Artificial Neural Network(ANN) analysis and see that compared to cut-based analysis, that parameter 
space can be probed with even lower luminosity.

Our work is organised as follows. In section~\ref{model} we present a brief outline of the model. In section~\ref{muonanomaly} we discuss the muon anomaly and its impact on our model parameter space. We then move to section~\ref{constraints} where we explore the allowed parameter space taking into account the theoretical constraints, constraints coming from low energy observables and from the collider.
We present a cut-based as well as neural-network-based collider analysis in section~\ref{collider}. We summarize our results and conclude in section~\ref{conclusion}.


\section{The Generalised Two Higgs Doublet Model }\label{model}

In this section we briefly discuss the model in consideration. We follow the convention as in~\cite{Primulando:2016eod}~\footnote{For general 2HDM review 
one should look into Ref~\cite{Branco:2011iw}.}. Two scalar doublets $\Phi_1$ and $\Phi_2$ with hyper charge 
$Y=1$~\footnote{$Q=T_3+\frac{Y}{2}$.} are present in this model. The most general scalar 
potential can be written as:
\begin{equation}
\begin{aligned}
	{\cal V}_{2HDM} &= M_{11}^2 (\Phi_1^{\dagger} \Phi_1^{\phantom{\dagger}}) + M_{22}^2 (\Phi_2^{\dagger} \Phi_2^{\phantom{\dagger}}) -  [M_{12}^2(\Phi_1^{\dagger} \Phi_2^{\phantom{\dagger}}) + \text{h.c.}]  \\
&+ \frac{1}{2} \lambda_1 (\Phi_1^{\dagger} \Phi_1^{\phantom{\dagger}})^2 + \frac{1}{2} \lambda_2 (\Phi_2^{\dagger} \Phi_2^{\phantom{\dagger}})^2 + \lambda_3 (\Phi_1^{\dagger} \Phi_1^{\phantom{\dagger}})(\Phi_2^{\dagger} \Phi_2^{\phantom{\dagger}}) + \lambda_4 (\Phi_1^{\dagger} \Phi_2^{\phantom{\dagger}})(\Phi_2^{\dagger} \Phi_1^{\phantom{\dagger}}) \\
&+ \{\frac{1}{2} \lambda_5 (\Phi_1^{\dagger} \Phi_2^{\phantom{\dagger}})^2 + [\lambda_6 (\Phi_1^{\dagger} \Phi_1^{\phantom{\dagger}}) + \lambda_7 (\Phi_2^{\dagger} \Phi_2^{\phantom{\dagger}})] (\Phi_1^{\dagger} \Phi_2^{\phantom{\dagger}}) + \text{h.c.} \}.
\end{aligned}
\end{equation}
where ${h.c.}$ denotes the $\rm{Hermitian~Conjugate}$ term.

In general, $M_{12}^2$, $\lambda_5$, $\lambda_6$ and $\lambda_7$ can be complex while 
the rest of the parameters are real.
However, in this work we assume CP is conserved, therefore
$M_{12}^2$, $\lambda_5$, $\lambda_6$ and $\lambda_7$ are taken to be real. We should mention here in the absence of $Z_2$ symmetry ($\Phi_1 \rightarrow \Phi_1, \Phi_2 \rightarrow - \Phi_2$) $\lambda_6$ and $\lambda_7$ can take non-zero values in general~\footnote{Throughout our paper we have varied $\lambda_6$ and $\lambda_7$ from 0 to 0.1.}.

The two scalar doublets  of the model can be expanded as 
\begin{equation}
	\Phi_1 = \begin{pmatrix}\phi_1^+\\ \frac{1}{\sqrt{2}}\left(Re[\Phi_1^0] + i Im[\Phi_1^0]\right)\end{pmatrix},
	\qquad
	\Phi_2 = \begin{pmatrix}\phi_2^+\\ \frac{1}{\sqrt{2}}\left(Re[\Phi_2^0] + i Im[\Phi_2^0]\right)\end{pmatrix},
\end{equation}
After electroweak symmetry breaking the doublets acquire vacuum expectation value(VEV).

\begin{equation}
\langle \Phi_1 \rangle=\frac{1}{\sqrt{2}} \left(
\begin{array}{c} 0\\ v_1\end{array}\right), \qquad \langle
\Phi_2\rangle=
\frac{1}{\sqrt{2}}\left(\begin{array}{c}0\\ v_2
\end{array}\right)\,.\label{potmin}
\end{equation}

A key parameter of the model is $\tan \beta = \frac{v_2}{v_1}$. Charged Goldstones($G^{\pm}$) and physical charged Higgs state ($H^{\pm}$) are produced as a linear combination of $\phi_1^{\pm}$ and $\phi_2^{\pm}$. On the other hand, the same linear combination of $Im[\Phi_1^0]$ and $Im[\Phi_2^0]$ gives rise to neutral CP-odd Goldstone ($G^0$) and physical CP-odd state ($A$). The mixing is shown in the following equations.

\begin{equation}
	\begin{pmatrix}\phi_1^{\pm}\\ \phi_2^{\pm}\end{pmatrix}= \begin{pmatrix} \phantom{-}\cos\beta & \sin\beta \\ -\sin\beta & \cos\beta\end{pmatrix}
	\begin{pmatrix}H^{\pm}\\ G^{\pm} \end{pmatrix},
\label{chargemixing}
\end{equation}

\begin{equation}
	\begin{pmatrix}\sqrt{2}Im[\Phi_1^0]\\ \sqrt{2}Im[\Phi_2^0]\end{pmatrix}= \begin{pmatrix} \phantom{-}\cos\beta & \sin\beta \\ -\sin\beta & \cos\beta\end{pmatrix}
	\begin{pmatrix}A\\ G^0 \end{pmatrix},
\label{pseudomixing}
\end{equation}




Diagonalizing the mass matrix for the CP-even neutral states we get the mass eigenstates $h$ and $H$, where the states in the mass basis and in the flavor basis are related by the following rotation. 
\begin{equation}
	\begin{pmatrix}\sqrt{2}Re[\Phi_1^0] - v_1\\ \sqrt{2}Re[\Phi_2^0] - v_2\end{pmatrix}= \begin{pmatrix} \phantom{-}\cos\alpha & \sin\alpha \\ -\sin\alpha & \cos\alpha\end{pmatrix}
	\begin{pmatrix}h\\ H \end{pmatrix},
\end{equation}

Where either $h$ or $H$ is assumed to behave like the Higgs of Standard Model with mass 125 GeV.

Having discussed the Higgs potential of generalized 2HDM, we proceed towards the Yukawa sector of the model. It is well-known that to avoid tree-level flavor-changing neutral current, a $Z_2$ symmetry is imposed on the general Yukawa Lagrangian of 2HDM. The doublets $\Phi_1$, $\Phi_2$ and the fermion fields behave either even or odd under this $Z_2$ symmetry and depending on this behavior four common types of 2HDM, namely Type I, Type II, Lepton-specific or Type X and Flipped or Type Y are formed. In Type I, up and down type quarks and leptons couple to $\Phi_2$. In Type II, up-type quarks couple to $\Phi_2$ and down-type quarks and leptons couple to $\Phi_1$. In Type X model up and down type quarks couple to $\Phi_2$ and leptons couple to $\Phi_1$. In Type Y 2HDM, up type quarks and leptons couple to $\Phi_2$ and the down-type quarks couple to $\Phi_1$.

Unlike these aforementioned types of 2HDM, in the generalized 2HDM, no $Z_2$ symmetry is imposed on the Lagrangian and therefore this model produces tree-level FCNC. In this case both the doublets $\Phi_1$ and $\Phi_2$ couple to all the leptons and 
quarks. In the generalized 2HDM, the Yukawa couplings to the quarks and leptons can be written as:

\begin{equation}
-{\cal L}_{Yukawa} = \bar{Q}_L (Y^d_1\Phi_1 + Y^d_2\Phi_2) d_R + \bar{Q}_L (Y^u_1 \tilde{\Phi}_1 + Y^u_2 \tilde{\Phi}_2) u_R + \bar{L}_L (Y^\ell_1\Phi_1 + Y^\ell_2\Phi_2) e_R + h.c.
\label{type3yuk}
\end{equation}

In Eq.~\ref{type3yuk}$, Y_{1,2}^{u,d,\ell}$ are the Yukawa matrices whose flavor indices have been suppressed and $\tilde{\Phi}_i = i\sigma_2\Phi_i^*$. Expanding this equation in terms of the VEVs and physical fields, we can get the fermion mass matrix.

\begin{equation}
{\bar f}_L {\bf M}^f f_R = {\bar f}_L (\frac{v_1Y^f_1}{\sqrt{2}} + \frac{v_2Y^f_2}{\sqrt{2}})  f_R + h.c.
\label{mass}
\end{equation}

Without assuming any particular relation between the matrices $Y_1$ and $Y_2$ it is not possible to diagonalize the two of them simultaneously, which leads to tree-level scalar mediated FCNC. To diagonalize the fermion mass matrices by bi-unitary transformation we need two unitary matrices $U_L^f$ and $U_R^f$. We adopt the convention of \cite{Crivellin:2015hha} and consider the Yukawa Lagrangian as a perturbation of Type X model in terms of FCNC couplings. Therefore we diagonalize $Y_2^u$, $Y_2^d$ and $Y_1^\ell$ matrices where $Y_1^u$, $Y_1^d$ and $Y_2^\ell$ remain non-diagonal leading to tree-level FCNC in the Yukawa sector. After diagonalization the Yukawa Lagrangian involving the neutral scalars can be written as follows.

 \begin{align}
 -{\cal L}^\phi_{Yukawa} & = \bar u_L \left[ \left( \frac{c_\alpha {\bf m}^u}{v s_\beta} - \frac{c_{\beta-\alpha} \Sigma^u}{\sqrt{2} s_\beta}\right) h + \left( \frac{s_\alpha {\bf m}^u}{ s_\beta v} + \frac{s_{\beta-\alpha} \Sigma^u}{\sqrt{2} s_\beta}\right) H \right] u_R \nonumber \\
&+ \bar d_L \left[ \left( \frac{c_\alpha {\bf m}^d}{v s_\beta} - \frac{c_{\beta-\alpha} \Sigma^d}{\sqrt{2} s_\beta}\right) h + \left( \frac{s_\alpha {\bf m}^d}{ s_\beta v} + \frac{s_{\beta-\alpha} \Sigma^d}{\sqrt{2} s_\beta}\right) H \right] d_R \nonumber \\
 &+ \bar e_L \left[ \left( - \frac{s_\alpha {\bf m}^\ell}{v c_\beta} + \frac{c_{\beta-\alpha} \Sigma^\ell}{\sqrt{2} c_\beta}\right)h + \left( \frac{c_\alpha {\bf m}^\ell}{ c_\beta v} - \frac{s_{\beta-\alpha} \Sigma^\ell}{\sqrt{2} c_\beta}\right) H \right] e_R \nonumber \\
 %
 %
& - i \left[ \bar u_L \left( \frac{ {\bf m}^u}{ t_\beta v} - \frac{ \Sigma^u }{\sqrt{2} s_\beta}\right) u_R 
 + \bar d_L \left(- \frac{ {\bf m}^d}{ t_\beta v} + \frac{ \Sigma^d }{\sqrt{2} s_\beta}\right) d_R 
 + \bar e_L \left( \frac{ t_\beta{\bf m}^\ell}{ v} - \frac{ \Sigma^\ell }{\sqrt{2} c_\beta}\right) e_R 
 \right] A + h.c.  \label{eq:Yu_phi}
 \end{align}

Here ${\bf m}^f$ = $U_L^f {\bf M}^f U_R^f$ is the diagonal mass matrices of the fermions and $U_L^f$ and $U_R^f$ are the unitary matrices required to diagonalize ${\bf M}^f$, $\Sigma^u = U_L^u Y_1^u U_R^{{\dagger}u}$, $\Sigma^d = U_L^d Y_1^d U_R^{{\dagger}d}$ and $\Sigma^\ell = U_L^\ell Y_2^u U_R^{{\dagger}l}$. 
$c_\alpha = \cos \alpha, s_\alpha = \sin \alpha, c_{\beta - \alpha} = \cos(\beta - \alpha), s_{\beta - \alpha} = \sin(\beta - \alpha) $ and $t_{\beta} = \tan \beta$.
It can be easily checked that the FCNC processes occur due to the presence of non-zero $\Sigma^{f}$ matrices. When $\Sigma^{f} = 0 $ the Yukawa couplings in Eq.~\ref{eq:Yu_phi} reduce to the couplings in the Type X 2HDM. Following the convention of \cite{Cheng:1987rs} $\Sigma^{f}$ matrices are parameterized in terms of dimensionless free parameters $\chi^f$s in the following manner. 

\begin{equation}
\Sigma^{f}_{ij} = \sqrt{m_i^f m_j^f} \chi^f_{ij}/v
\label{chi}
\end{equation}

In general $\chi^f_{ij}$ may not be equal to $\chi^f_{ji}$, but for simplicity we assume $\chi^f_{ij}$ = $\chi^f_{ji}$ in our analysis. We mention here that the non-diagonal couplings of the pseudoscalar $A$ (see Eq.~\ref{eq:Yu_phi}) will be crucial for our study and we will call them $y_{\mu e}$, $y_{\tau e}$ and $y_{\mu\tau}$ in the following sections.

As the rotation matrix for charged scalars is the same as 
that  for pseudoscalars which can be seen from Eq.~(\ref{chargemixing} and \ref{pseudomixing}), therefore, the  Yukawa couplings of the charged Higgs boson are similar to those of the CP-odd scalar and can be written as

\begin{align}
{\cal L}^{H^\pm}_Y & = \frac{\sqrt{2}}{v} \bar u_{i }  \left( m^u_i  (\xi^{u*})_{ki}  V_{kj} P_L + V_{ik}  (\xi^d)_{kj}  m^d_j P_R \right) d_{j}  H^+  \nonumber \\
& + \frac{\sqrt{2}}{v}  \bar \nu_i  (\xi^\ell)_{ij} m^\ell_j P_R \ell_j H^+ + h.c. \label{eq:Yukawa_CH}
\end{align}

Where the sum over flavor indices is indicated, $V\equiv U^u_L U^{d\dagger}_L$ is the Cabibbo-Kobayashi-Maskawa (CKM) matrix, and $P_{R,L}=(1 \pm \gamma_5)/2$ are the chiral projection operators. The expressions for $\xi^{f}$ matrices are given below. 

\begin{eqnarray}
\xi^{u} = \frac{1}{t_{\beta}} \delta_{ij} - \frac{1}{\sqrt{2}s_{\beta}} \sqrt{\frac{m_i^u}{m_j^u}} \chi_{ij}^u, \\
\xi^{d} = -\frac{1}{t_{\beta}} \delta_{ij} + \frac{1}{\sqrt{2}s_{\beta}} \sqrt{\frac{m_i^d}{m_j^d}} \chi_{ij}^d, \\
\xi^{\ell} = t_{\beta} \delta_{ij} - \frac{1}{\sqrt{2}c_{\beta}} \sqrt{\frac{m_i^\ell}{m_j^\ell}} \chi_{ij}^\ell
\end{eqnarray}

Here also we can see that non-zero $\Sigma$ matrices and equivalently non-zero $\xi$ matrices will introduce non-trivial coupling structure even in the charged Higgs interaction with the quarks and leptons. One can check that in the absence of these matrices the couplings reduce to couplings in the Type X 2HDM.


\section{Exploration of Muon ($g-2$)}
\label{muonanomaly}

The muon anomalous magnetic moment is one of the biggest triumphs of quantum field theory. A precise measurement of this helps one to comprehend the higher order corrections contributing to it. Moreover, it indicates an existence of new physics because of the long-standing discrepancy between SM prediction and experimental observation~\cite{Blum:2013xva}. Possibly the ongoing E989 experiment at Fermilab~\cite{Grange:2015fou} and future E34 experiment at J-PARC~\cite{Iinuma:2011zz} will shed new light on this discrepancy between the theory and experiment.

In classical quantum mechanics the value of $g_{\mu}$(gyromagnetic ratio for $\mu$) is 2. It receives correction from loop effects in quantum field theory. These corrections are parameterized in terms of $a_{\mu} = \frac{g_{\mu}-2}{2}$. In SM it receives contribution via QED, electroweak and hadronic loops. A great deal of effort has been put forth to  determine the SM prediction to an unprecedented level of accuracy. SM contributions up to three orders in the electromagnetic constant, has been calculated by~\cite{Davier:2010nc,Hagiwara:2011af,Davier:2017zfy,Davier:2019can}. Taking into account pure QED, electroweak and hadronic contribution, the predicted value for muon anomaly in the SM is given by\cite{Aoyama:2020ynm,Davier:2017zfy,Keshavarzi:2018mgv,Colangelo:2018mtw,Hoferichter:2019mqg,Davier:2019can,Keshavarzi:2019abf,Kurz:2014wya,Melnikov:2003xd,Masjuan:2017tvw,Colangelo:2017fiz,Hoferichter:2018kwz,Gerardin:2019vio,Bijnens:2019ghy,Colangelo:2019uex,Colangelo:2014qya,Blum:2019ugy,Aoyama:2012wk,Czarnecki:2002nt,Aoyama:2019ryr,Gnendiger:2013pva}
\begin{equation}
a_{\mu}^{SM}= 116591810(43)  \times 10^{-11}
\end{equation}
The most recent bound comes from BNL(2006) data~\cite{Bennett:2006fi} which gives
\begin{equation}
a_{\mu}^{exp} = 116592089(63)  \times 10^{-11}
\end{equation}

The difference between the theory and experiment denotes a $3.7\sigma$ discrepancy which can certainly leave us a room to search for NP scenarios.
\begin{equation}
\Delta a_{\mu}=  a_{\mu}^{exp} - a_{\mu}^{SM} = 279(76)  \times 10^{-11}
\label{mgm2}
\end{equation}

\begin{figure}[!hptb]
	\centering
	\includegraphics[width=14cm,height=5.0cm]{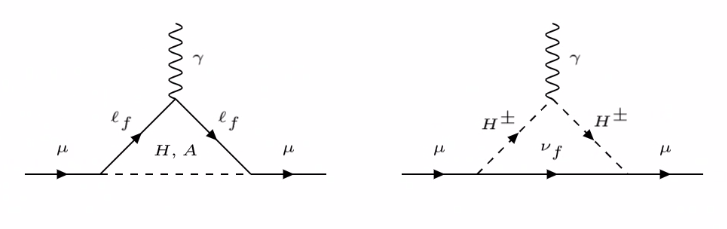}

	\caption{\it Non-standard contribution to $\Delta a_{\mu}$ at one-loop. }
	
	\label{mgm21loop}
\end{figure}

\begin{figure}[!h]
	\centering
	\includegraphics[width=12cm,height=11cm]{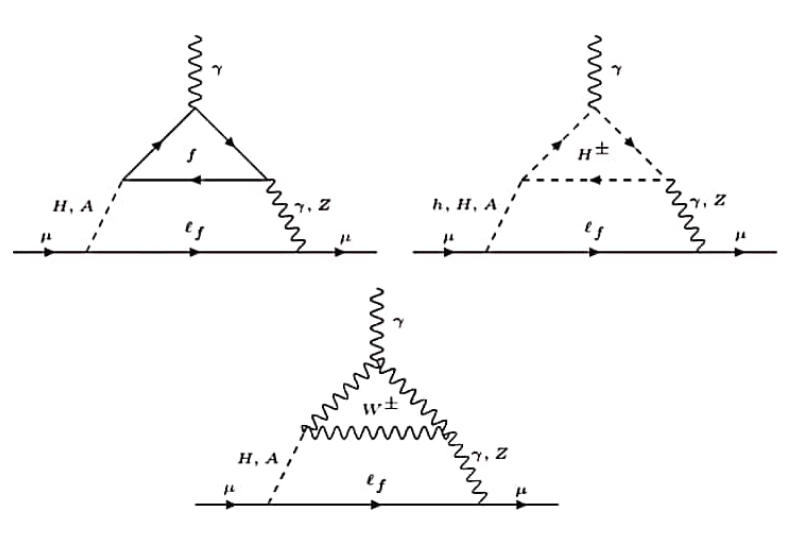}

	\caption{\it Non-standard contribution to $\Delta a_{\mu}$ from two-loop Bar-Zee diagrams with internal $\gamma/Z$. }
	
	\label{intgamma}
\end{figure}

\begin{figure}[!hptb]
	\centering
	\includegraphics[width=14cm,height=11cm]{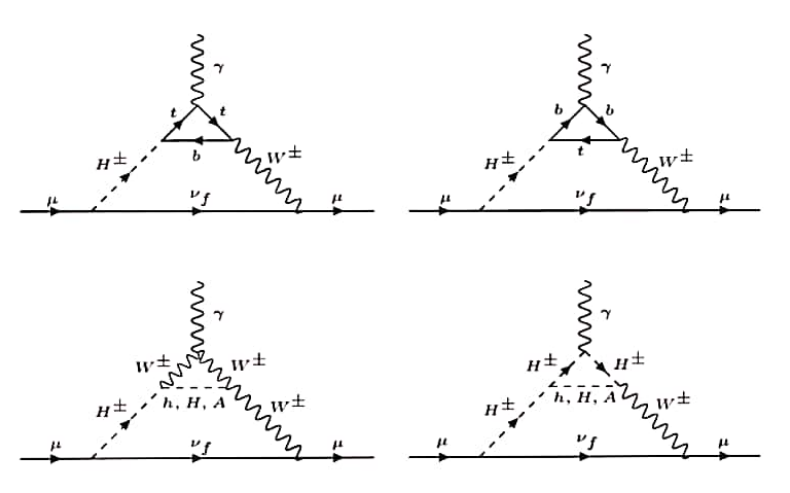}
	\caption{\it Non-standard contribution to $\Delta a_{\mu}$ from two-loop Bar-Zee diagrams with internal $W^{\pm}$ and $H^{\pm}$. }
	
	\label{barzeewhp}
\end{figure}

\begin{figure}[!hptb]
	\centering
	\includegraphics[width=15cm,height=5cm]{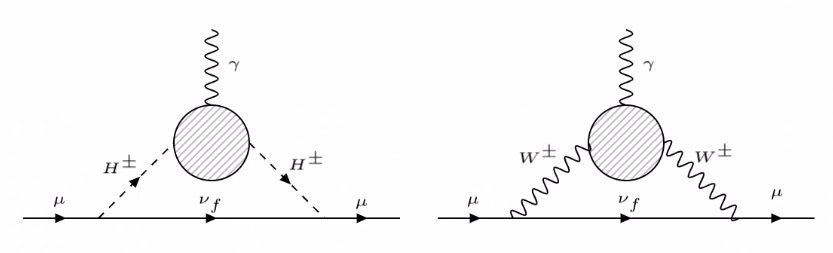}
	\caption{\it Non-standard contribution to $\Delta a_{\mu}$ from two-loop Bar-Zee diagrams same as Fig.~\ref{barzeewhp}, with internal $W^{\pm}$ or $H^{\pm}$. }
	
	\label{barzeeall}
\end{figure}

In this work we consider one loop as well as two loop Bar-Zee type contribution to $\Delta a_{\mu}$ in generalized 2HDM. It is shown in earlier works~\cite{Queiroz:2014zfa,Ilisie:2015tra} that the two-loop Bar-Zee diagrams can bring sizeable contribution for a large region of parameter space. We present the diagrams for one loop contribution to $\Delta a_{\mu}$ in Fig.~\ref{mgm21loop} and two loop Bar-Zee diagrams in Fig.~\ref{intgamma},~\ref{barzeewhp},~\ref{barzeeall}. We mention here that the two loop Bar-Zee contributions dominate over the one-loop contributions. Although the two loop diagrams have a loop suppression factor but also have an enhancement factor of $\frac{M^2}{m_{\mu}^2}$, where $M$ is the mass of the heavy particle running in the loop namely $t, b, H^{\pm}, W^{\pm}$ as can be seen from Fig.~\ref{intgamma}. This enhancement factor usually dominates over the loop suppression. The Bar-Zee contribution from an internal photon and heavy fermion or $H^{\pm}$ running in the loop has been studied in great detail in the past and these diagrams give rise to major contribution to $\Delta a_{\mu}$. The contribution from diagrams where a $Z$ boson replaces the internal photon is negligible due to coupling as well as mass suppression. Also, the diagram involving $W^{\pm}$ in the loop, will have negligible contribution because of suppression in the coupling between $W^{\pm}$ bosons and non-standard CP-even Higgs in the alignment limit. We have also considered the Bar-Zee diagrams where charged Higgs replaces the neutral Higgs and $W^{\pm}$ substitutes the internal $\gamma$ in Fig.~\ref{barzeewhp}. It has been shown in \cite{Ilisie:2015tra} that their contribution can be sizeable in some regions of the parameter space. In Fig.~\ref{barzeeall}, we also consider the diagrams with internal charged Higgs or $W^{\pm}$ where the grey circle represents the same loops as in Fig.~\ref{barzeewhp}, excluding the fermion loops for the $W^{\pm}$ diagram, because that will be a pure SM contribution.

\begin{figure}[!hptb]
\centering
\includegraphics[width=80mm,height=59.0mm]{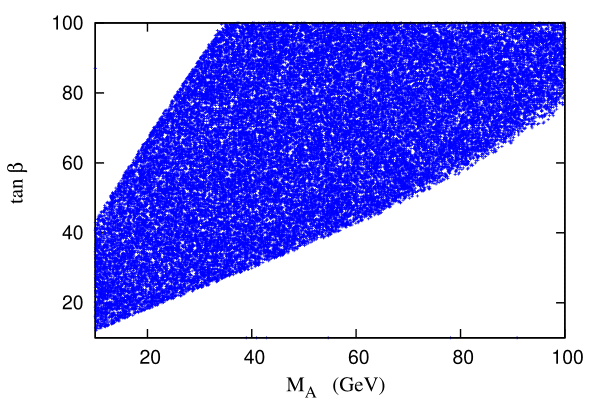}
\caption{\it The allowed region in $m_A - \tan \beta$ plane from $g_{\mu}-2$ data at $3 \sigma$. The flavor changing couplings are taken to be $y_{\mu e}= 10^{-7}, y_{\tau e}=5\times 10^{-5}, y_{\mu \tau}=5\times 10^{-5}$. The non-standard neutral CP-even Higgs mass is 120 GeV and charged Higgs mass is 150 GeV.}
\label{muon_anomaly}
\end{figure}

We compute $\Delta a_{\mu}$ taking into account all the aforementioned diagrams following~\cite{Queiroz:2014zfa,Ilisie:2015tra}. Next we scan the parameter space of our model and plot allowed region in the $m_A - \tan \beta$ plane in Fig.~\ref{muon_anomaly}. For the scanning, the flavor changing couplings are taken to be $y_{\mu e}= 10^{-7}, y_{\tau e}=5\times 10^{-5}, y_{\mu \tau}=5\times 10^{-5}$. The choice of such Yukawa couplings will be discussed shortly in the next section. The non-standard neutral CP-even Higgs mass and charged Higgs mass are fixed at 120 GeV and 150 GeV. We mention here that the parameter space allowed by $\Delta a_{\mu}$ has been explored in the context of Type X 2HDM~\cite{Broggio:2014mna,Cherchiglia:2017uwv}. We have considered the most updated experimental bound, exhaustive set of one and two-loop diagrams and also the effect of lepton-flavor violating vertices, as compared to the earlier works. One can check that a low mass pseudoscalar with an enhanced coupling to the $\tau$ leptons will give significant contribution to $\Delta a_{\mu}$(see Fig.~\ref{intgamma}(top left)). In our model the coupling of pseudoscalar with a pair of $\tau$ leptons is proportional to $\tan \beta$. Therefore, low $m_A$ and large $\tan\beta$ region is favored in the light of $g_{\mu}-2$ data. While scanning the parameter space we have used the $3\sigma$ upper and lower bound on the experimentally observed central value of $\Delta a_{\mu}$(Eq.~\ref{mgm2}).

\section{Constraints on the model}\label{constraints}

We have seen from the discussion in the previous section that the major contribution to the anomalous magnetic moment of muon comes from the 
low mass pseudoscalar contribution at moderate to large $\tan \beta$. However, in the presence of the non-diagonal Yukawa matrices it is inevitable that the similar contributions will also generate FCNC processes. This flavor changing processes include loop induced $\mu \rightarrow e \gamma$, $\tau \rightarrow e \gamma$, $\tau \rightarrow \mu \gamma$, $\mu \rightarrow 3e$ and $\mu-e$ conversion, all of which put a strong constraint on the flavor changing couplings as well as the (pseudo)scalar masses and $\tan \beta$. It is evident that low mass pseudoscalar and large $\tan \beta$ regions will be disfavored from the limits coming from the low energy measurements, which seems to be in tension with the requirement of muon ($g-2$). We study these constraints carefully in the upcoming subsection and explore the regions of parameter space consistent with the limits from the non-observation of low energy flavor violating processes as well as the experimental observation of $(g_{\mu}-2)$. Moreover, these flavor changing vertices also give rice to FCNC in the (pseudo)scalar mediated tree-level decays in the leptonic final states. Our main objective of this work is to probe this region of parameter space in the collider. We mention here that to get sufficient event rate at the collider, we focus on the low mass range of the decaying (pseudo)scalar. We now proceed to discuss various constraints on our model which further guide us to choose our benchmarks for the upcoming direct search analysis at the collider.

\subsection{Limits from low energy measurements}

 In the SM, lepton flavor is conserved since neutrinos are massless. In neutrino oscillation~\cite{Fukuda:1998mi,Ahmad:2002jz}, LFV has been observed in the neutrino sector. 
However, till date LFV has not been observed in the charged lepton sector. Therefore, lepton flavor violation can be treated as one of the important tools to search for new physics. 
Many new physics models can accommodate LFV processes. Since, no such signal has been observed yet, there are strong limits on these LFV processes~\cite{Aubert:2009ag}. We will soon see that in the low $m_A$ region, which is of our primary interest from the $(g_{\mu}-2)$ requirements, the LFV processes will also be dominated by the pseudoscalar contribution in the loop. Therefore these limits from the low energy LFV processes will essentially constrain the non-diagonal lepton Yukawa couplings of the pseudoscalar $A$, namely $y_{\mu e}$, $y_{\tau e}$ and $y_{\mu\tau}$.

The recent bound on $ BR(\mu \to e \gamma) < 4.2 \times 10^{-13}$ comes from MEG experiment~\cite{TheMEG:2016wtm}. 
 The other important constraint in LFV will come from $\mu \to 3 e$ which is a natural consequence of $\mu \to  e \gamma$ decay when the resulting photon converts to $e^+e^-$ pair. 
Apart from that, $\mu - e$ conversion in nuclei can also be an important signature of LFV.
Assuming that the chirality flipping dipole term dominates, the relation between $\mu \to e \gamma$ and other possible LFV constraints, namely, $\mu - e$ conversion(CR) and $BR(\mu \to 3 e)$ can be roughly estimated as ~\cite{Kuno:1999jp,Lindner:2016bgg} 
\begin{subequations}
\begin{eqnarray}
CR(\mu~ Ti \to e~ Ti) \simeq \frac{1}{200} BR(\mu \to e \gamma) \\
BR(\mu \to 3 e) \simeq \frac{1}{160} BR(\mu \to e \gamma)
\end{eqnarray}
\label{conv}
\end{subequations}

We emphasize that the relations quoted in Eq.~\ref{conv} are model-dependent~\cite{Crivellin:2014cta} and here we merely try to give a rough order of magnitude estimation between the observables. 
If we try to translate the limits according to the relations in Eq.~\ref{conv}, we find that the limit on $CR(\mu~ Ti \to e~ Ti)$  has to be $ < 2.1 \times 10^{-15}$ for it to be of same strength as the limit from $\mu \rightarrow e \gamma$, while from experiment~\cite{Bartolotta:2017mff} this upper limit is $6.1 \times 10^{-13}$~\cite{Wintz:1998rp}. 
Similar argument holds for $BR(\mu \to 3 e)$ branching ratio. From  Eq.~\ref{conv}, $BR(\mu \to 3 e)$ should be $< 2.62 \times 10^{-15}$, for this upper limit to be of same strength as BR($\mu \rightarrow e \gamma$) while the experimental upper bound is $1.0 \times 10^{-12}$~\cite{Bellgardt:1988qe}. %
Therefore, it is evident that these two constraints are relatively weak. Hence, for our analysis we take into account the strongest bound which is coming from $\mu \to e \gamma$
\footnote{However, just as a passing comment we would like to highlight the fact that the next generation of experiments such
as Mu2e and COMET will use Aluminum as target aiming at a sensitivity of
$\sim 10^{-17}$ on $\mu-e$ conversion~\cite{Bernstein:2013hba} and probably then it would be the strongest constraint among all the LFV observables.}.

Similar to the $\mu-e$ sector, there are strong constraints on $(\tau \to e \gamma)$ and $(\tau \to \mu \gamma)$  branching ratio.
Current Bound on $ BR(\tau \to e \gamma) < 3.3 \times 10^{-8}$~\cite{Aubert:2009ag} and  $ BR(\tau \to \mu \gamma) < 4.4  \times 10^{-8}$~\cite{Aubert:2009ag} puts a strong constraint on $y_{\tau e}$ and $y_{\mu\tau}$ respectively. One should also take into account the limits on the BR($\tau \rightarrow 3e) < 2.7 \times 10^{-8}$~\cite{Beringer:1900zz} and BR($\tau \rightarrow 3\mu) < 2.1 \times 10^{-8}$~\cite{Beringer:1900zz}. However, compared to $BR(\tau \to e \gamma)$, the limit on BR($\tau \rightarrow 3e$) is weaker due to an addition suppression of factor $\alpha$.
Same is true for the limit on BR($\tau \rightarrow 3\mu)$ which puts much weaker limit compared to $ BR(\tau \to \mu \gamma)$.
 
\begin{figure}[!hptb]
\centering
\includegraphics[width=14cm,height=5.0cm]{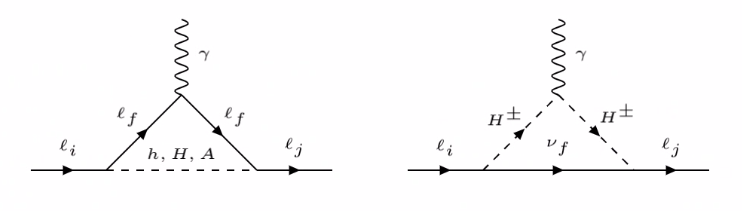}
\caption{\it One loop contribution to lepton flavor violating decays.}
\label{1loop_fcnc}
\end{figure}

We calculate the LFV processes at one and two-loop following references~\cite{Lindner:2016bgg,Omura:2015xcg}. The presence of flavor non-diagonal Yukawa matrices in the generalized 2HDM gives rise to flavor violating coupling between scalars and leptons at the tree-level. This in turn enables the LFV decays at one as well as two-loop. We show the corresponding one-loop diagrams in Fig.~\ref{1loop_fcnc}. One can see that these diagrams are same as Fig.~\ref{mgm21loop} with modification in the incoming and outgoing fermion lines. Likewise, the two loop diagrams taking part in LFV decays can be obtained from Figs.~\ref{intgamma} to \ref{barzeeall} by similar modifications. It is worth mentioning that the loop contribution from the neutral scalars dominates over the contribution from the charged scalar loop (see Fig.~\ref{1loop_fcnc}). We have found that the two loop contribution to $\tau \rightarrow \mu \gamma$ and $\tau \rightarrow e \gamma$ amplitudes add up to mere $\sim 2\%$ of their one-loop counterpart. On the contrary, in case of $\mu \rightarrow e \gamma$, the addition of two loop contribution induces 3 times enhancement to the one-loop amplitude. 

It is interesting to note that BR($\tau \rightarrow e \gamma$) and BR($\tau \rightarrow \mu \gamma$) constrain the couplings $y_{\tau e}$ and $y_{\mu\tau}$ respectively. The reason behind this is the following. The major contribution to the corresponding amplitudes come from the $\tau$ loop (see Fig.~\ref{1loop_fcnc} (left)), where $y_{\tau e}$ and $y_{\mu \tau }$ appear with $m_{\tau}$. Therefore these terms dominate over the other terms which are accompanied by $m_{\mu}$, $m_{e}$ or are product of two LFV couplings. Hence the upper limit on the aforementioned branching ratios  constrains particularly $y_{\tau e}$ and $y_{\mu\tau}$. However, the situation is different in case of BR($\mu \rightarrow e \gamma$). In this case the $\tau$ loop has the highest contribution in terms of loop integral. The $\tau$-loop integral at one loop comes with a coefficient $y_{\mu\tau} \times y_{\tau e}$. Hence its contribution can be comparable with the $e$ or $\mu$ loop, with coefficients $y_{\mu e}$ multiplied with $m_e$ or $m_{\mu}$. Therefore BR($\mu \rightarrow e \gamma$) is not solely controlled by $y_{\mu e}$ at one loop. However at two loop it is the  $y_{\mu e}$ coupling that dominates the amplitude.

We have seen that BR($\tau \rightarrow e\gamma$) constrains $y_{\tau e} < 10^{-4}$ and BR($\tau \rightarrow \mu\gamma$ ) constrains
 $y_{\tau \mu} < 10^{-4}$ . However, for $y_{\mu e}$ the situation is not so straightforward. Unlike $\tau \rightarrow e \gamma$ and $\tau \rightarrow \mu \gamma$, the decay $\mu \rightarrow e \gamma$ does not primarily constrain $y_{\mu e}$ coupling as discussed earlier. However, calculating the amplitudes at two-loop, to satisfy all the three LFV conditions simultaneously along with muon anomaly, the coupling $y_{\mu e}$ gets a strong upper bound ($ < 10^{-6}$).

\begin{figure}[!h]
	\centering
	\includegraphics[width=7cm,height=6cm]{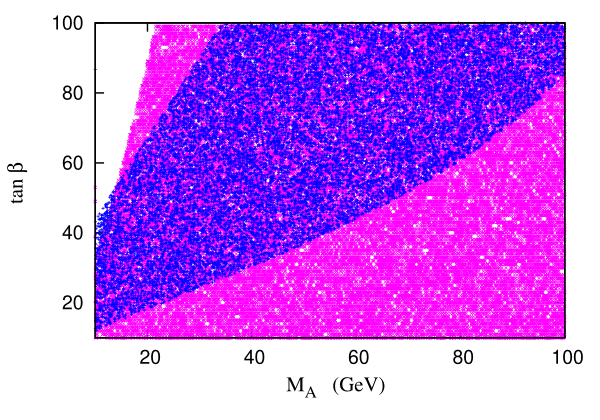}
	\includegraphics[width=7cm,height=6cm]{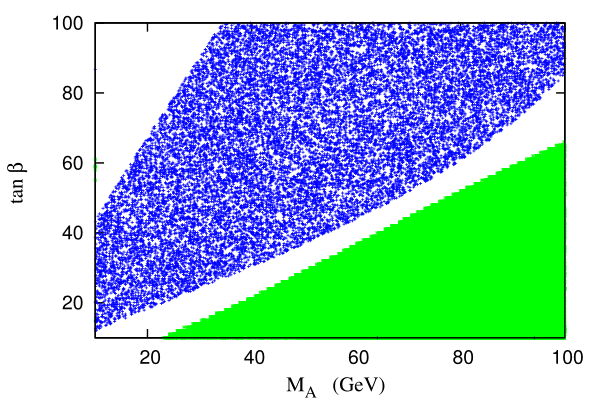}
	\includegraphics[width=7cm,height=6cm]{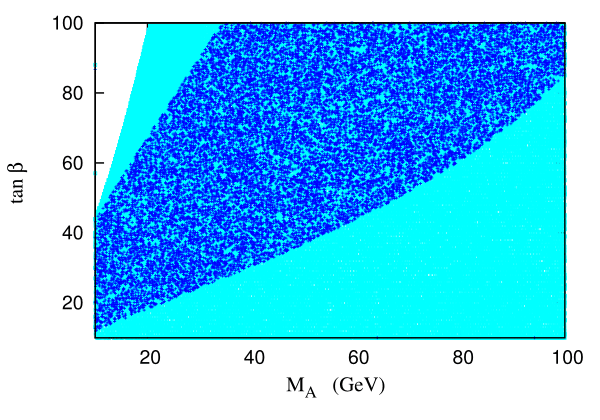}
	\caption{\it The magenta, green and cyan regions are the allowed range for $\mu \to e \gamma$,
	$\tau \to e \gamma$ and $\tau \to \mu \gamma$ respectively. The blue band is the 
	3$\sigma$ allowed range for muon anomaly. 
	The flavor changing couplings are taken to be $y_{\mu e}= 10^{-7},
	y_{\tau e}= 10^{-4},y_{\mu \tau}= 10^{-5}$. The non-standard neutral CP-even Higgs mass is 120 GeV and charged Higgs mass is 150 GeV.}
	
	\label{case1}
\end{figure}

\begin{figure}[!h]
	\centering
	\includegraphics[width=7cm,height=6cm]{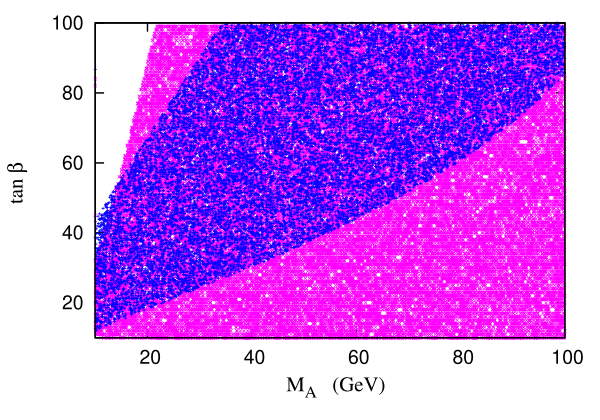}
	\includegraphics[width=7cm,height=6cm]{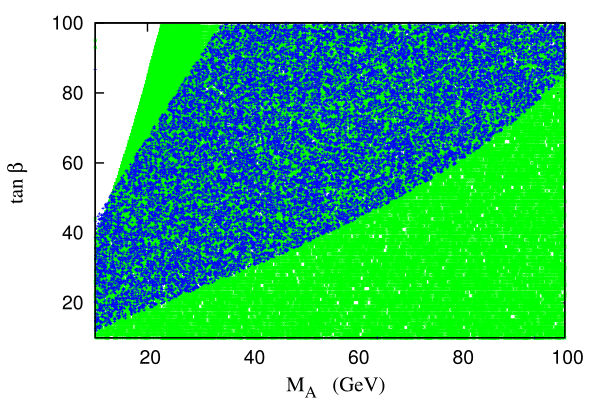}
	\includegraphics[width=7cm,height=6cm]{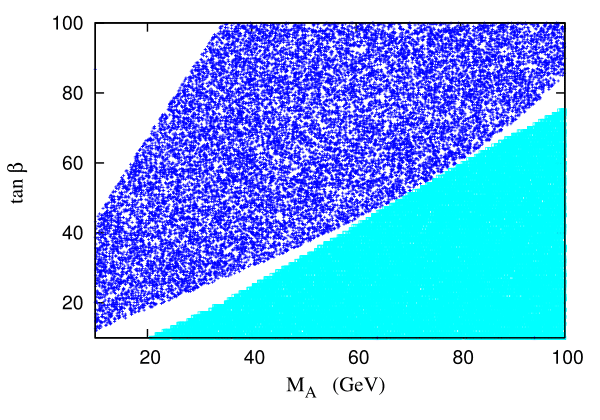}
	\caption{\it The magenta, green and cyan regions are the allowed range for $\mu \to e \gamma$,
	$\tau \to e \gamma$ and $\tau \to \mu \gamma$ respectively. The blue band is the  
	3$\sigma$ allowed range for muon anomaly. 
	The flavor changing couplings are taken to be $y_{\mu e}= 10^{-7},
	y_{\tau e}= 10^{-5},y_{\mu \tau}= 10^{-4}$. The non-standard neutral CP-even Higgs mass is 120 GeV and charged Higgs mass is 150 GeV.}
	
	\label{case2}
\end{figure}

\begin{figure}[!h]
	\centering
	\includegraphics[width=7cm,height=6cm]{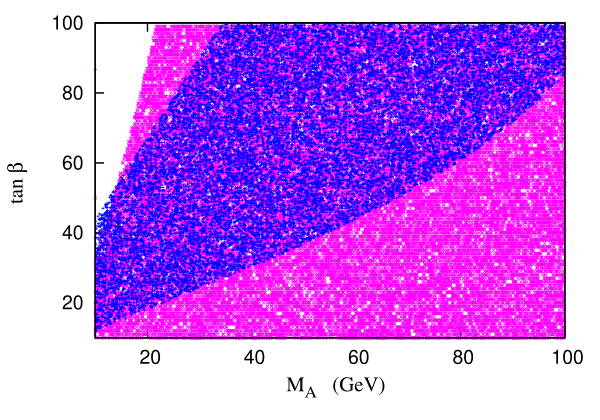}
	\includegraphics[width=7cm,height=6cm]{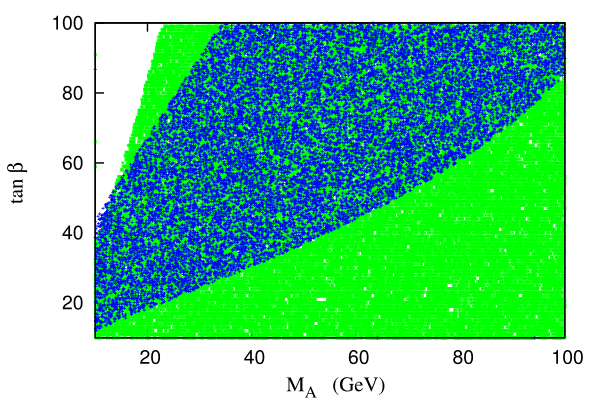}
	\includegraphics[width=7cm,height=6cm]{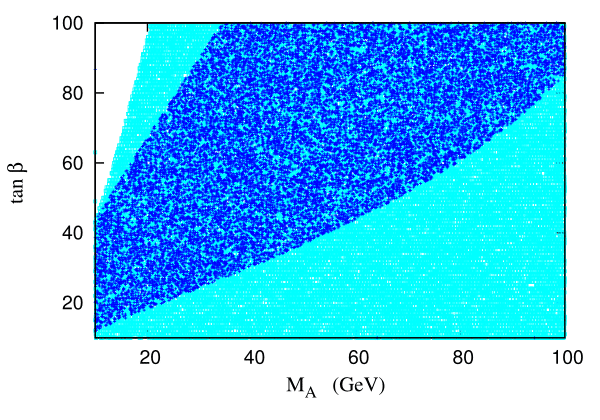}
	\caption{\it The magenta, green and cyan regions are the allowed range for $\mu \to e \gamma$,
	$\tau \to e \gamma$ and $\tau \to \mu \gamma$ respectively. The blue band is the allowed 
	3$\sigma$ allowed range for muon anomaly. 
	The flavor changing couplings are taken to be $y_{\mu e}= 10^{-7},
	y_{\tau e}= 10^{-5},y_{\mu \tau}= 10^{-5}$. The non-standard neutral CP-even Higgs mass is 120 GeV and charged Higgs mass is 150 GeV.}
	
	\label{case3}
\end{figure}

\begin{figure}[!h]
	\centering
	\includegraphics[width=7cm,height=6cm]{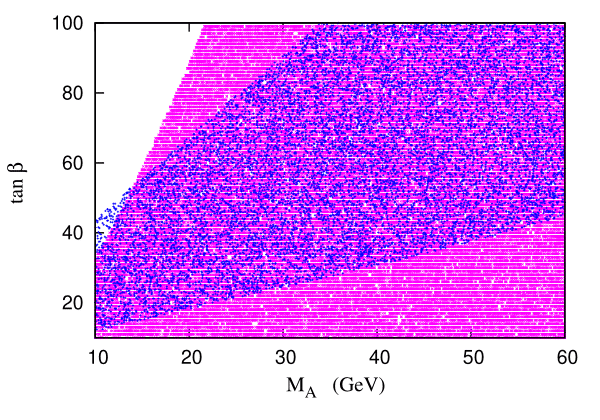}
	\includegraphics[width=7cm,height=6cm]{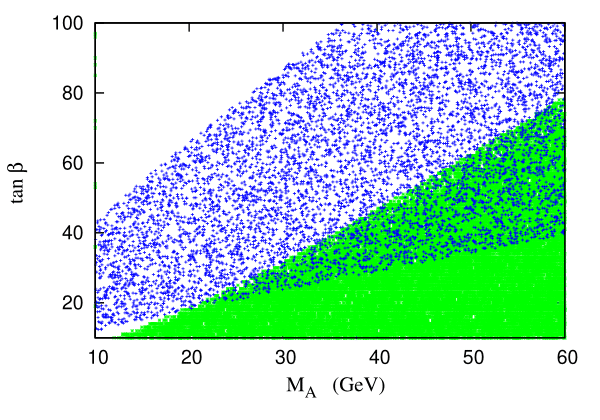}
	\includegraphics[width=7cm,height=6cm]{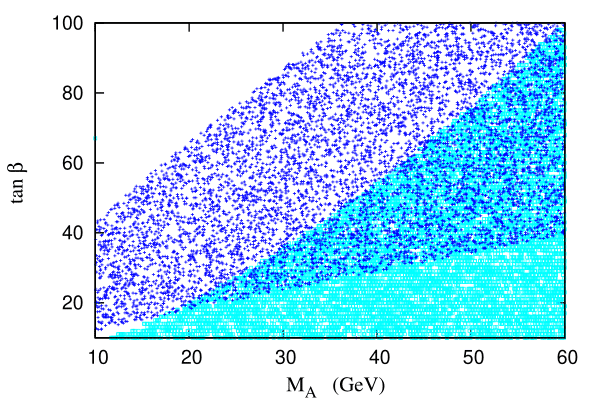}
	\caption{\it The magenta, green and cyan regions are the allowed range for $\mu \to e \gamma$,
	$\tau \to e \gamma$ and $\tau \to \mu \gamma$ respectively. The blue band is the allowed 
	3$\sigma$ allowed range for muon anomaly. The overlapping regions satisfy both constraints.
	The flavor changing couplings are taken to be $y_{\mu e}= 10^{-7},
	y_{\tau e}=5\times 10^{-5},y_{\mu \tau}=5\times 10^{-5}$. The non-standard neutral CP-even Higgs mass is 120 GeV and charged Higgs mass is 150 GeV.}
	
	\label{case4}
\end{figure}

It is important to note that the branching ratios we just described also depend strongly on the scalar masses and $\tan \beta$, along with the flavor changing couplings. In Fig.~\ref{case1}-\ref{case4} we have plotted the regions allowed by LFV constraints in $m_A - \tan \beta$ plane for specific choices of flavor changing Yukawa couplings. In Fig.~\ref{case1}-\ref{case4}, we have also superimposed the region allowed by $(g_{\mu}-2)$ data on the region allowed by low energy LFV data.

It can be clear from Fig~\ref{case1},\ref{case2} that the two limits tend to favor non-overlapping regions, unless the LFV Yukawa couplings are below certain values. In Fig.~\ref{case3} and ~\ref{case4}, we present our choice of LFV Yukawa couplings for which we get the an overlapping region that is allowed by $(g_{\mu}-2)$ as well as low energy LFV constraints. We specifically concentrate on the scenario depicted in Fig.~\ref{case4}, because the values of flavor violating Yukawa couplings ($y_{\mu e}= 10^{-7}, y_{\tau e}=5\times 10^{-5},y_{\mu \tau}=5\times 10^{-5}$) in this case produce adequate event rate which can be probed at the HL-LHC. Therefore this region is of primary interest to us from the collider point of view.

\subsection{Theoretical constraints}
\label{theory_constraints}

The constraints from the requirements of vacuum stability and perturbativity have been studied in detail in earlier works~\cite{Broggio:2014mna,Gunion:2002zf}. It has been pointed out that large separation between the $m_A$ and $m_H^{\pm}$ is disfavored from the theoretical considerations of vacuum stability and perturbativity. Since we are interested in the low mass pseudoscalars from the requirements of $(g_{\mu}-2)$, it is imperative to check the upper limit on $m_H^{\pm}$ compatible with low $m_A$. The vacuum stability and perturbativity conditions put bounds on the $\lambda$ parameters and thereby correlate the masses of different neutral and charged scalars.  The vacuum stability condition requires~\cite{Deshpande:1977rw,Gunion:2002zf}

\begin{eqnarray}
\lambda_1 > 0,~~~~\lambda_2 > 0,~~~~\lambda_3 > -\sqrt{\lambda_1 \lambda_2},~~~~\lambda_3+\lambda_4-|\lambda_5| > \sqrt{\lambda_1 \lambda_2}
\end{eqnarray}

The resulting squared-masses for the CP-odd and charged Higgs states are given by~\cite{Gunion:2002zf}

\begin{eqnarray}
m_A^2 &=& \frac{m_{12}^2}{s_\beta c_\beta} - \frac{1}{2}v^2(2\lambda_5 + \frac{\lambda_6}{t_\beta} + \lambda_7 t_\beta) \\
m_{H^{\pm}}^2 &=& m_A^2 + \frac{1}{2} v^2 (\lambda_5 - \lambda_4)
\label{massdiff}
\end{eqnarray}

It is clear from Eq.~\ref{massdiff}, the mass-square difference ${m_H^{\pm}}^2 - m_A^2$ is proportional to $\lambda_5 - \lambda_4$, which should be less than $\lambda_3 + \sqrt{\lambda_1\lambda_2}$. Along with the vacuum stability criteria, the requirement of perturbativity, ie. all the quartic couplings $C_{H_iH_jH_kH_l} < 4\pi$ puts an upper limit on $m_H^{\pm} - m_A$. In the scenario when $m_h = 125$ GeV, the parameter space allowed by stability and perturbativity constraints are indicated in \cite{Broggio:2014mna}. We have performed a scan in the following range of parameters for the  scenario where the mass of the heavier CP-even neutral Higgs $m_H = 125$ GeV (The justification behind this choice will be discussed shortly) and hard $Z_2$-symmetry breaking parameters $\lambda_6$ and $\lambda_7$ are assumed to be non-zero. We have assumed alignment limit in the analysis and therefore have varied the mixing angle $\cos(\beta - \alpha)$ close to unity. 

 $m_A \in$ [10.0 GeV, 60.0 GeV],~~$m_H \in$ [62.5 GeV, 125.0 GeV],~~$m_H^{\pm} \in$ [89.0 GeV, 190.0 GeV],~~$m_{12}^2 \in$ [-1000 GeV$^2$, 1000 GeV$^2$]$,~~\tan \beta \in$ [10, 70],~~$|\cos(\beta - \alpha)| \in$ [0.99, 1],~~$\lambda_6 \in$ [0, 0.1],~~$\lambda_7 \in$ [0, 0.1]

\begin{figure}[!hptb]
\centering
\includegraphics[width=7cm, height=6cm]{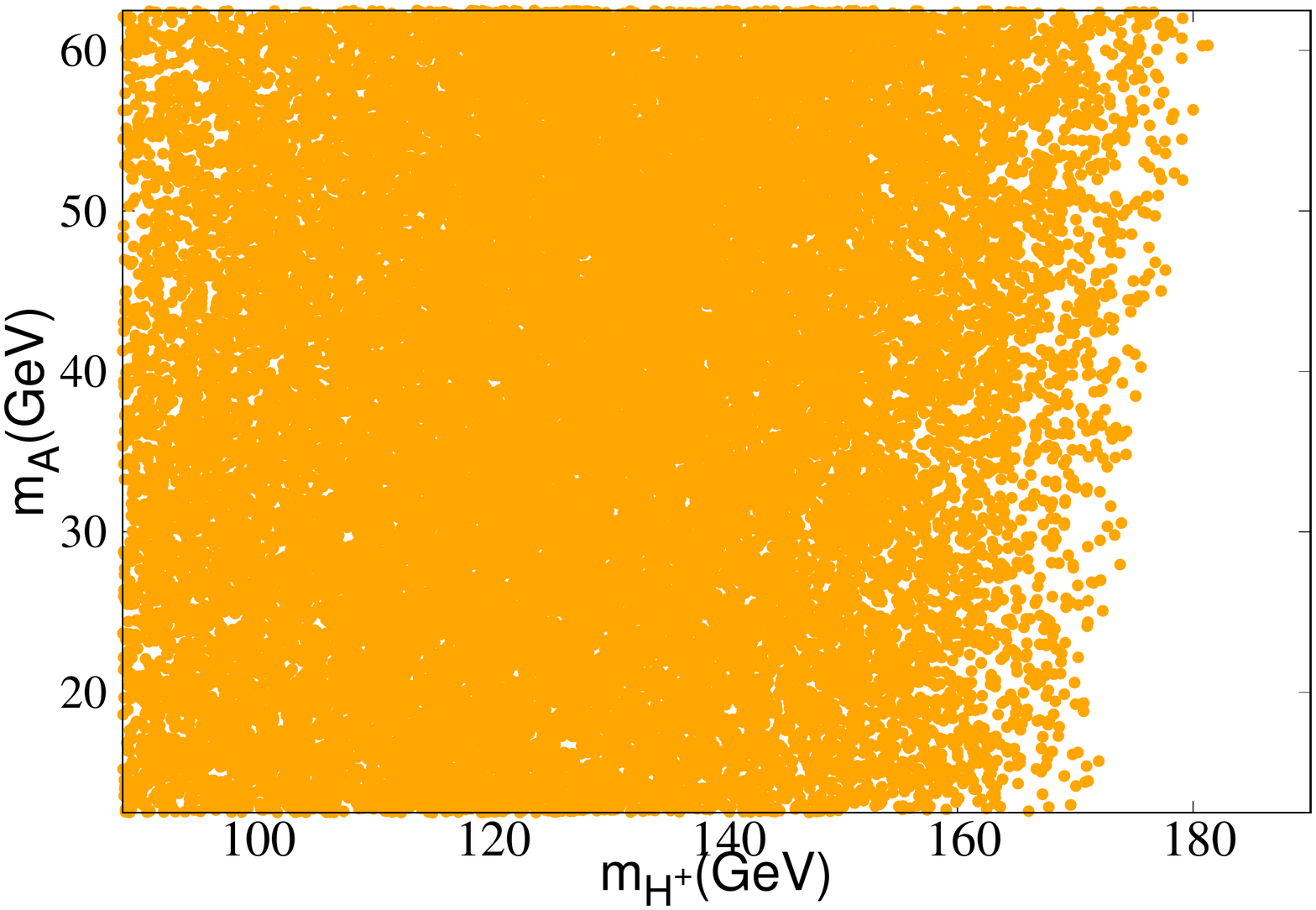} 
\includegraphics[width=7cm, height=6cm]{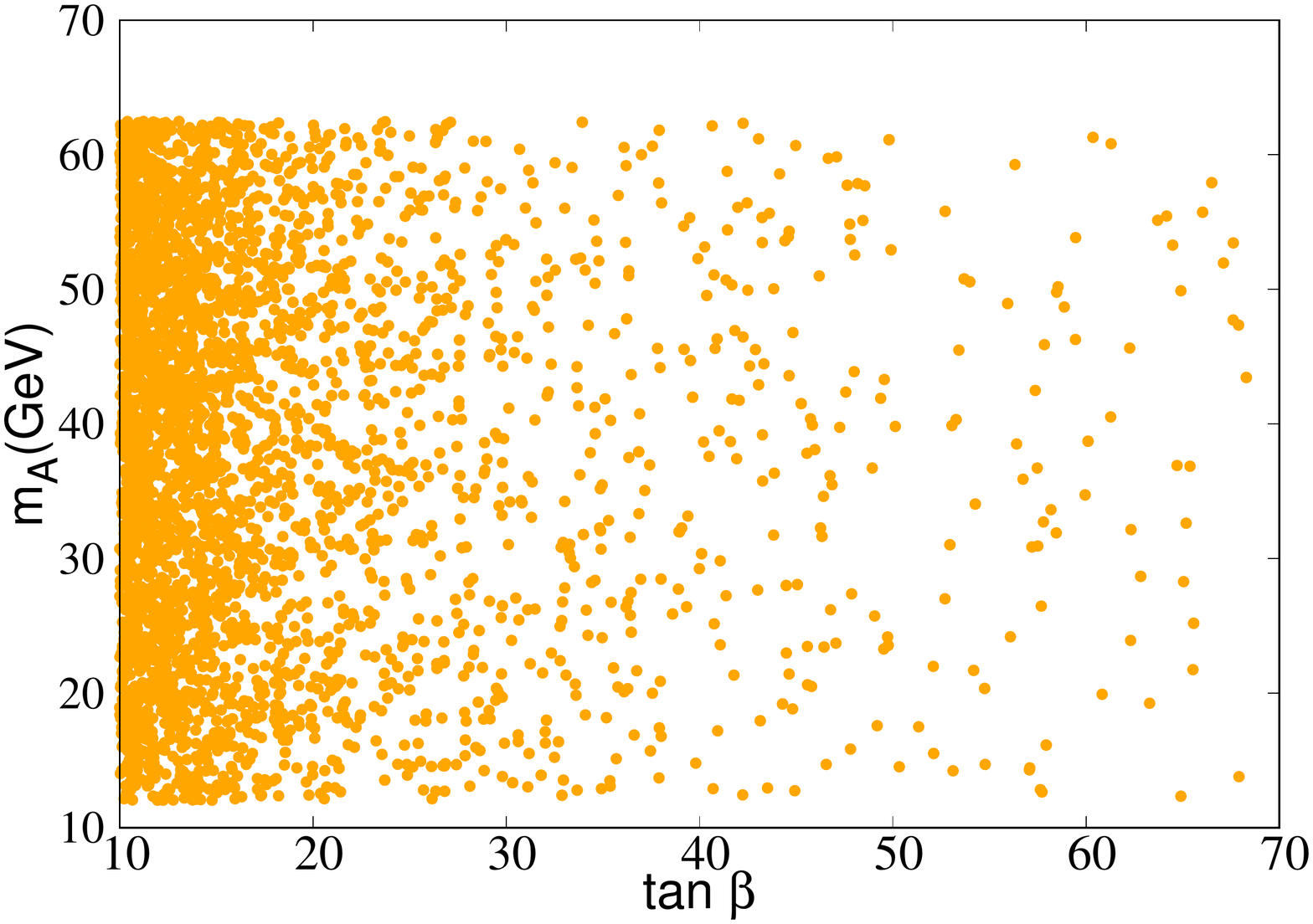} 
\caption{\it Parameter space allowed by stability, unitarity and perturbativity constraints.}
\label{constraints_theory}
\end{figure}

In Fig.~\ref{constraints_theory} we show the parameter space allowed by stability, unitarity and perturbativity constraints. We show only
the low $m_A$ region as we will be interested in this region in our collider analysis. In Fig.~\ref{constraints_theory}(left) we show the
upper limit of $m_{H^{\pm}}$ as a function of $m_A$ as pointed out in the foregoing discussion. We can see that $m_{H^{\pm}} < 170-180$ GeV
is allowed for low $m_A$. In Fig.~\ref{constraints_theory}(right) we show that constraints in the $\tan\beta-m_A$ plane. We see that although very large
$\tan \beta$ is allowed from perturbativity considerations, low to moderate $\tan \beta$ values are much more favored compared to the high values.

Using the relations between the quartic couplings $\lambda$'s and the physical masses and Higgs mixing parameter $m_{12}^2$, one can find the $hAA$ coupling~\cite{Gunion:2002zf,Cherchiglia:2017uwv}

\begin{eqnarray}
g_{hAA} = \frac{1}{2v}\left[(2m_A^2 - m_h^2) \frac{\cos(\alpha-3\beta)}{\sin 2\beta} + (8 m_{12}^2 - \sin 2\beta(2m_A^2 + 3m_h^2)) \frac{\cos(\beta+\alpha)}{\sin^2 2\beta}\right]  \nonumber \\
 + v\left[\sin 2\beta \cos 2\beta(\lambda_6-\lambda_7)\sin(\beta - \alpha) - (\lambda_6 \sin\beta \sin 3\beta + \lambda_7 \cos \beta \cos 3\beta)\cos(\beta - \alpha)\right] 
\end{eqnarray}

It is important to notice that the low $m_A$ region of parameter space which we are interested in, yield a substantial branching fraction for $h \rightarrow AA$ decay, where $h$ is the 125 GeV Higgs and $m_A < \frac{m_h}{2}$. The experimental upper limit on this branching ratio is rather strong~\cite{Sirunyan:2020eum}, where a stringent limit comes from the search for ($p p \rightarrow h \rightarrow AA)$ process in the $\mu^+\mu^-\tau^+\tau^-$ final state. The only way such a small branching ratio can be achieved is when the coupling $g_{hAA}$ is extremely small. This in turn imposes a relation between $m_{12}^2$, $\tan\beta$ and $m_A$~\cite{Bernon:2014nxa}. However $m_{12}^2$ is a parameter a crucial parameter which ensures perturbativity. It is required for perturbativity  that $m_{12}^2 \sim \frac{m_H^2}{\tan\beta}$. It is shown in~\cite{Bernon:2014nxa}, in the case where 125 GeV Higgs is the lightest Higgs boson, and $m_H > 125$ GeV, it is possible to obey the perturbativity constraints as well as the upper limit on BR($h \rightarrow AA)$ for low $\tan \beta < 10$ and the mass gap $m_H - m_h$ is not very large. Although this region is phenomenologically viable, the ($g_{\mu}-2)$ requirements(see Fig.~\ref{muon_anomaly}) impose that $m_A$ should also be very small, ie $m_A < 10$ GeV. This statement is only valid in the `right-sign' region of 2HDM where Higgs coupling with the fermions and gauge bosons are of same sign. In the so-called `wrong-sign' region where Higgs coupling to the fermions and gauge bosons are of opposite sign, gives rise to entirely different allowed region and phenomenological signatures. We have not explored this scenario in the current work and will be pursuing in a future study.

The other possibility is to consider the case when the heavier CP even Higgs is SM-like, ie $m_H = 125$ GeV. However in this case the LEP limit implies either $m_A$ or $m_h$ can be $< \frac{m_H}{2}$. We consider the low mass pseudoscalar, and therefore $m_h > \frac{m_H}{2}$.
Here also, like the previous case the limit on BR($h \rightarrow AA$) will indicate extremely small value of the coupling $g_{HAA}$ whose expression is given as follows: 

\begin{eqnarray}
g_{HAA} = \frac{1}{2v}\left[(2m_A^2 - m_H^2) \frac{\cos(\alpha-3\beta)}{\sin 2\beta} + (8 m_{12}^2 - \sin 2\beta(2m_A^2 + 3m_H^2)) \frac{\cos(\beta+\alpha)}{\sin^2 2\beta}\right] \nonumber \\
+ v\left[\sin 2\beta \cos 2\beta(\lambda_6-\lambda_7)\cos(\beta - \alpha) + (\lambda_6 \sin\beta \sin 3\beta + \lambda_7 \cos \beta \cos 3\beta)\sin(\beta - \alpha)\right] 
\end{eqnarray}

In this case there is more freedom compared to the previous case, in terms of the allowed parameter space. One can have a pseudoscalar mass $>10$ GeV with moderate $\tan \beta$, with suitable value of $m_{12}^2$ and $m_h$, while satisfying perturbativity condition and the small BR($H \rightarrow AA$) simultaneously. This point onwards, we will explore this particular scenario, ie. for our work $m_H = 125$ GeV.

\subsection{Electroweak constraints}

In  this  section  we  analyze  the  impact of constraints arising from electroweak precision measurements, especially the oblique parameters~\cite{Lavoura:1993nq,Baak:2012kk} on our model. The experimental collaboration Gfitter group~\cite{BAAK:2014gga} has published a contour in the plane of $S$ and $T$ parameter taking into account the correlation between them. The status of 2HDM in the light of the most recent global electroweak data has been presented in~\cite{Haller:2018nnx}. We mention here that we have used the elliptic contour which has been computed with $U$ as a free parameter. This choice leaves us with a less constrained parameter space compared to the cases $U=0$.

\begin{figure}[!hptb]
	\centering
	\includegraphics[width=8cm,height=7cm]{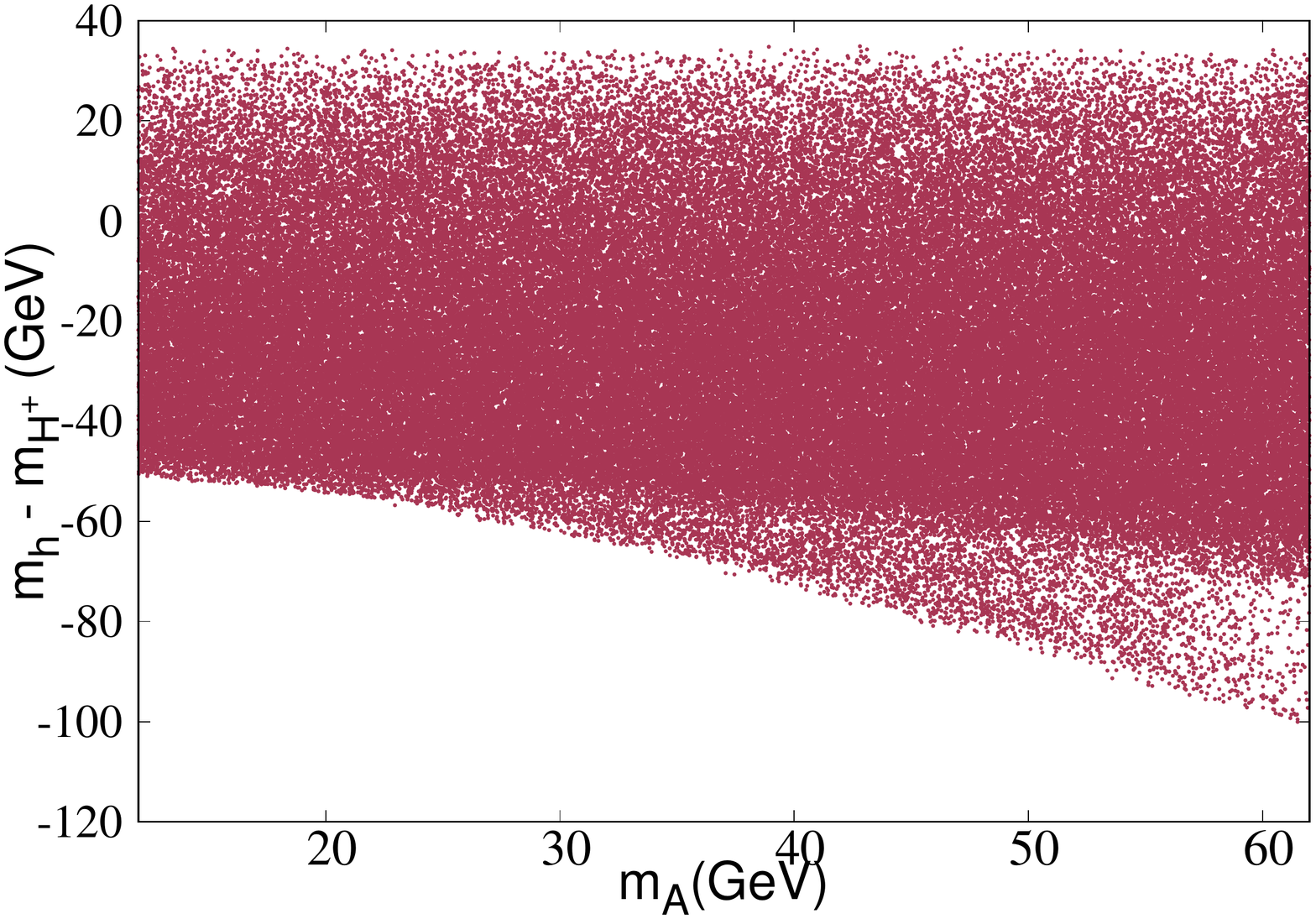}
	\caption{\it Parameter space satisfying electroweak constraints in the plane of $m_A$ and $m_h - m_{H^{\pm}}$.}
	\label{stu}
\end{figure}

We have calculated the oblique parameters in our model and obtained the allowed region of parameter space at 3$\sigma$. Here also we concentrate on low $m_A$ region and considered the case $m_H = 125$ GeV, ie. the second lightest CP-even Higgs is SM-like. $m_{H^{\pm}}$ has been varied from 90 GeV - 200 GeV. In Fig.~\ref{stu}, We present the allowed region in the plane of $m_A$ and $(m_h - m_{H^{\pm}})$. It is evident from the figure that as the pseudoscalar mass decreases, the mass difference $m_h - m_{H^{\pm}}$ should also decrease to obey the constraints from oblique parameters. We would like to point out that we have chosen the scalar masses $m_h = 120$ GeV, $m_{H^{\pm}} = 150$ GeV, (where the neutral CP even heavier Higgs mass $m_H$ is 125 GeV) as representative points in earlier part of our analysis and also for collider studies. These choices are governed by the requirement of simultaneous satisfaction of the theoretical as well as electroweak constraints.

\subsection{Constraints from B-physics}

From the discussions of Section~\ref{model}, it is clear that in presence of flavor changing couplings in the Yukawa sector the charged Higgs couplings to quarks and leptons are also modified. With new free parameters in the Lagrangian, interesting phenomenologies are likely to show up in rare decay processes that were suppressed in the SM. One such possibility in the rare processes involving $B-$meson. The free parameters of the model gets constrained by the experimental upper bounds on such rare FCNC processes. It is clear from Eq.~\ref{chi} that the FCNC within the first two generations are naturally suppressed by the small quark masses, while a larger freedom is allowed for the FCNC in the top and bottom quark sector. In our analysis also we have taken only $\lambda_{tt}$ and $\lambda_{bb}$ to be non-zero where $\lambda_{tt}$ and $\lambda_{bb}$ are the $h t \bar t$ and $h b \bar b$ coupling strengths respectively, considering $h$ to be the non-SM like CP-even Higgs. 

The rare processes involving $B$-mesons primarily include the decay $B \rightarrow X_s \gamma$, $B_s \rightarrow \mu^+ \mu^-$, $B^{\pm} \rightarrow \tau^{\pm} \nu_{\tau}$, $B^0-\bar{B}^0$ mixing whose strength is determined by the mass difference $\Delta M_B$ between the two states. The most stringent constraint comes from the $B \rightarrow X_s \gamma$ decay. The impact of these constraints in terms of specific types of 2HDM as well as in generalised 2HDM have been studied in great detail in earlier works~\cite{Crivellin:2013wna,Arbey:2017gmh,Hussain:2017tdf}. In conventional type I and type II 2HDM, the dominant additional contribution to the loop induced decay $B \rightarrow X_s \gamma$ comes from the charged Higgs boson-top quark penguin diagrams and its contribution depends on $m_{H^{\pm}}$. In type II 2HDM, this extra contribution interferes constructively with its SM counterpart and therefore the lower bound on the charged Higgs boson becomes rather high ($m_H^{\pm} \gtrsim 600$ GeV). In type I, the charged Higgs penguin diagram's contribution interferes destructively with its SM counterpart and gives negligible result at large $\tan \beta$. Therefore no strong constraint appear on the mass of the charged Higgs in type I model. The type X model has same structure as type I, as far as the interactions of Higgs with the quark sector is concerned. Therefore Type X models also do not receive any strong lower bound on $m_{H^{\pm}}$. As we can think of our model as a perturbation from the type X scenario, in the absence of the extra terms in the Yukawa Lagrangian, there is no strict lower bound on the charged Higgs mass. However, even in the presence of non-zero FCNC Yukawa matrix elements, it is possible to have low enough charged Higgs mass~\cite{Xiao:2003ya,Mahmoudi:2009zx,Arhrib:2017yby,Cherchiglia:2017uwv,Enomoto:2015wbn} with suitable choice of $\lambda_{tt}$ and $\lambda_{bb}$ couplings. We have taken in our analysis $\lambda_{tt} \sim 0.5$ and $\lambda_{bb} \sim 2$, which allows a charged Higgs mass $m_{H^{\pm}} \gtrsim 150$ GeV.   

Another decay process which can constrain our model parameters space is $B^{\pm} \rightarrow \tau^{\pm} \nu_{\tau}$ where charged Higgs enters at the tree level itself. The observed branching ratio for the process $B_u^{\pm} \rightarrow \tau^{\pm} \nu_{\tau} = (1.06 \pm 0.19) \times 10^{-4}$~\cite{Alonso:2016oyd}. The decay $B_c^{\pm} \rightarrow \tau^{\pm} \nu_{\tau}$, although has not been observed, but puts an upper limit ($< 30\%$)~\cite{Alonso:2016oyd} on the branching ratio for this decay. However, we have assumed only $\lambda_{tt}$ and $\lambda_{bb}$ are non-zero in the quark sector, we find out that these limits essentially reduces to a limit on $\lambda_{bb}$ and $\tan \beta$. In~\cite{Arhrib:2017yby}, it has been shown that $\lambda_{bb} \sim 2$ is favored for large or moderate $\tan \beta$. 

The constraint from $\Delta M_B$ puts an an upper limit on $\lambda_{tt}$ as a function of the charged Higgs mass~\cite{Mahmoudi:2009zx}. $m_{H^{\pm}} \gtrsim 150$ GeV is allowed for $\lambda_{tt} \lesssim 0.5$. Therefore our choice of parameter space obeys this constraint as well.

The upper limit on the BR($B_s \rightarrow \mu^+ \mu^-$) is 2.4$^{+0.9}_{-0.7} \times 10^{-9}$~\cite{Patrignani:2016xqp}. This particular branching fraction constrains the low $\tan \beta (< 2$) region for low $m_H^{\pm}(\sim 100$ GeV)~\cite{Arbey:2017gmh}. For higher charged Higgs mass this limit is further relaxed.

\subsection{Constraints from direct search at the colliders}

Constraints can be obtained from collider searches for the production and decay of on-shell neutral and charged Higgs bosons. There have been numerous searches in the past in this direction. The LEP experiments have looked for pair production of charged Higgs bosons in the process $e^+ e^- \rightarrow \gamma/Z \rightarrow H^+ H^-$. In this process all the couplings that appear are essentially gauge couplings, this predictions for this process therefore depends only on the charged Higgs mass $m_{H^{\pm}}$. However the decay and branching fractions of $H^{\pm}$ are indeed model dependent. But a combined search for $H^{\pm}$ in $\tau \nu$ and $c \bar s$ channel put a robust  lower limit of 80 GeV on $m_{H^{\pm}}$~\cite{Abbiendi:2013hk}. This limit only mildly depends on BR($H^{\pm} \rightarrow \tau \nu)$. 

At the LHC the charged Higgs search can be categorized in two types. For $m_H^{\pm} < m_t$, charged Higgs can be produced from the decay of top quark($t \rightarrow bH^{\pm})$. This decay has been searched for in $\tau \nu$~\cite{Aad:2014kga,Khachatryan:2015qxa} and $c\bar s$~\cite{Aad:2013hla,Khachatryan:2015uua} final state. These searches have put an upper limit on BR($t \rightarrow bH^{\pm})\times(H^{\pm}\rightarrow \tau \nu/c \bar s)$. The other important search mode at the LHC is $(p p \rightarrow tbH^{\pm}$) in the final states $\tau \nu$~\cite{Khachatryan:2015qxa,Aaboud:2016dig} and $c \bar s$~\cite{ATLAS-CONF-2016-088,CMS-PAS-HIG-16-031} and $t \bar b$~\cite{ATLAS:2016qiq}.  

Collider searches for the non-standard neutral Higgs also put constraints on the parameter space of interest. Searches for non-standard  Higgs bosons are performed at the LHC in various channels with SM final states. As we are specifically interested in the low pseudoscalar mass region with enhanced coupling to leptons, the limits which are crucial for our case comes from the search for low pseudoscalar produced in association with $b$ quarks and decaying into $\tau \tau$ final state~\cite{Khachatryan:2015baw,CMS:2019hvr}. Constraints from the search for low mass (pseudo)scalar produced in association with $b \bar b$ and decaying into $b \bar b$~\cite{Khachatryan:2015tra,CMS:2016ncz} has been taken into account.
CMS has also searched for decay involving two non-standard Higgs bosons such as $h/H \rightarrow Z(\ell\ell)A(\tau\tau)$~\cite{Khachatryan:2016are} and $h/H \rightarrow Z(\ell\ell)A(b \bar b)$~\cite{CMS-PAS-HIG-16-010}. However these limits become applicable for heavier CP-even Higgs $\gtrsim 200$ GeV. Therefore these particular searches do not have any considerable affect on our parameter space. 

We mention here that one should also take into account the limits coming from the direct search of the 125 GeV Higgs in various final states
including $\tau \tau$~\cite{Aad:2015vsa,Chatrchyan:2014nva},$\mu \mu$~\cite{Aad:2014xva,Khachatryan:2014aep}. Moreover, as the focus of our work is FCNC in the Yukawa sector, the constraints coming from flavor violating decays of 125 GeV Higgs boson also put constraints on the flavor-violating Yukawa matrix elements. The 125 GeV Higgs decaying to $e\mu$ and $e\tau$ final state have been looked for by the CMS experiments~\cite{CMS:2015udp}. CMS also puts an upper limit on the branching ratio for 125 GeV Higgs decaying to $\mu \tau$ final state~\cite{Khachatryan:2015kon}. Undoubtedly, these limits are crucial for our study. However, as we strictly confine ourselves to alignment limit ($\cos(\beta - \alpha) \approx 0.999)$, the flavor violating decays of the 125 GeV Higgs($H$ in our case) will receive a suppression by a factor $\sin^2(\beta-\alpha$) which can be seen from Eq.~\ref{eq:Yu_phi}. Therefore in this limit the constraints coming from lepton flavor violating decays of the 125 GeV Higgs are trivially satisfied.

An important constraint comes from the direct search for 125 GeV Higgs decaying into two light pseudoscalar final states when it is kinematically allowed. The upper bound on this branching ratio translates into severe constraint on the parameter space of this model. We have discussed this in detail in a previous subsection~\ref{theory_constraints} and have taken into account in our analysis.


\section{Collider Searches}\label{collider}

From our discussions in the previous sections it is clear that the existence of flavor violation in the lepton Yukawa sector gives rise to flavor-violating decays of $\mu$ and $\tau$ leptons. The presence of off-diagonal elements in the Yukawa matrices are the source of the lepton flavor violation in generalized 2HDM. The flavor violating decays of leptons are introduced at loop level via flavor violating coupling between the scalar and leptons at tree level. These flavor non-diagonal tree-level Yukawa coupling between the scalar and leptons will also give rise to interesting phenomenology at the colliders~\cite{Banerjee:2016foh,Primulando:2016eod,Primulando:2019ydt,Jana:2020pxx}.

In this work we consider probing the CP-odd scalar $A$ in flavor violating leptonic 
decay mode in generalized 2HDM at the HL-LHC. Our signal process is given as

\begin{equation}
 p p \to A \to \ell \tau_{\ell'}
\end{equation}

\noindent
where $\ell, \ell'= e,\mu$ and $\tau_{\ell'}$ denotes the leptonic decay of $\tau$. The signal of our interest is $\ell^{+}\ell'^{-} + \slashed{E_T}$. 

The SM processes that can give rise to similar final states are $\tau\tau/ e e / \mu\mu ,  t \bar{t}, W^{\pm}$+jets,
di-boson, SM Higgs~\cite{Banerjee:2016foh, Sirunyan:2019shc}. Out of these backgrounds, the major background in our signal region is the 
$\tau\tau$. Due to large cross-section, $t\bar{t}$  also serves an important background. In reality, $t \bar t$ leptonic final state turns out to be an irreducible background, whereas $t \bar t$ semileptonic and $W+$ jets background, despite having significant cross section, reduce to a large extent by our preselection criteria, which will be discussed shortly. From now on, we indicate $t \bar t$ leptonic channel as $t \bar t$ background unless specified otherwise. The $ee/\mu\mu$ background also has considerable cross section. However, in our signal region, this background contributes only $<$ 5\% of the $\tau \tau$ background. Therefore we do not discuss this background explicitly. The di-boson and SM Higgs background has much smaller cross section compared to the aforementioned backgrounds and they turn out to be less severe.

For our analysis, we choose three benchmark points valid from all the experimental and theoretical constraints and quote their 
production cross-section in Table~\ref{tab:bp}. We mention here that since the branching ratios of the pseudoscalar decaying to flavor violating final states is very small (BR($A \to \mu \tau) \approx$ BR($A \rightarrow \tau e) \approx 10^{-7}$), owing to the smallness of lepton flavor violating Yukawa couplings, we are compelled to choose low mass pseudoscalar which will have considerable production cross-section and therefore will be a viable candidate to search for in the collider.   

\begin{table}[!hptb]
\centering
 \begin{tabular}{|c|c|c|c|c|c|c|c|c|c|}
  \hline
  & $\tan\beta$ &  $m_{A}$ & $m_h$ & $m_H^{\pm}$ & $m_{12}^2$ & $\lambda _6$ & $\lambda_7$ & $|\cos(\beta-\alpha)|$ &$\sigma_{prod}(\sqrt{s}=14 ~\rm{TeV})$\\
  &  &  (in GeV) & (in GeV) & (in GeV) & (in GeV$^2$) & & & & (in fb) \\
  \hline
  BP1 & 15 & 21 & 120 & 150 & 970 & 0.001 & 0.001 & 0.999 & 0.085 \\ \hline
  BP2 & 20 & 25 & 120 & 150 & 843 & 0.1 & 0.005 & 0.999 & 0.067 \\  \hline
  BP3 & 22 & 27 & 120 & 150 & 775 & 0.01 & 0.0045 & 0.999 & 0.052 \\ \hline   
  \hline
 \end{tabular}
	\caption{\it Benchmark points allowed by all constraints and the corresponding production cross-section of
	our signal at LO at 14 TeV LHC. }
	\label{tab:bp}
 \end{table}

We first present the cut-based analysis for this channel in the following subsection. Next we explore the possible improvement of our results with multivariate techniques using Artificial Neural Network (ANN).

\subsection{Cut-based Analysis}

The signal events are generated in {\tt Madgraph5@NLO}~\cite{Alwall:2014hca} implementing the model file in {\tt FeynRules}~\cite{Alloul:2013bka}. We generate both signal and SM backgrounds events at leading order (LO) 
in {\tt Madgraph5@NLO}~\cite{Alwall:2014hca} using the {\tt NNPDF3.0} parton distributions~\cite{Ball:2014uwa}. 
The parton showering and hadronization are done using the built-in {\tt Pythia}
\cite{Sjostrand:2006za} within Madgraph. 
The showered events are then passed through {\tt Delphes}(v3)~\cite{deFavereau:2013fsa} 
for the detector simulation where the jets are constructed 
using the anti-$K_{T}$ jet algorithm with minimum jet formation radius $\Delta R = 0.5$. 
The isolated leptons are considered to be separated from the jets and other leptons by $\Delta R_{\ell i} \gtrsim 0.4, i =j,\ell$.

To generate our signal and background events, we employ the following pre-selection cuts.
\begin{eqnarray}
 p_T(j,b) &>& 20 ~{\rm GeV}\,; \quad |\eta(j)| < 4.7 \,; \quad |\eta(b)| < 2.5 \,, \nonumber \\ 
 p_T(\ell) &>& 10 ~{\rm  GeV}\,, \quad |\eta(\ell)| <2.5 \,.
 \label{basic_cut}
\end{eqnarray}

The $b$-jets are tagged with the $p_T$-dependent $b$-tag efficiency following the criteria of Ref.~\cite{Sirunyan:2017ezt}
which has an average 75\% tagging efficiency of the $b$-jets with $50~{\rm GeV} < p_T < 200~{\rm GeV}$  and 1\% mis-tagging 
efficiency for light jets.

Additionally, we propose the following selection cuts on certain kinematic observables to disentangle the signal from the SM backgrounds that would enhance the signal significance. We describe those observable in the following.

\begin{figure}[!hptb]
	\centering
	\includegraphics[width=7.2cm,height=6.0cm]{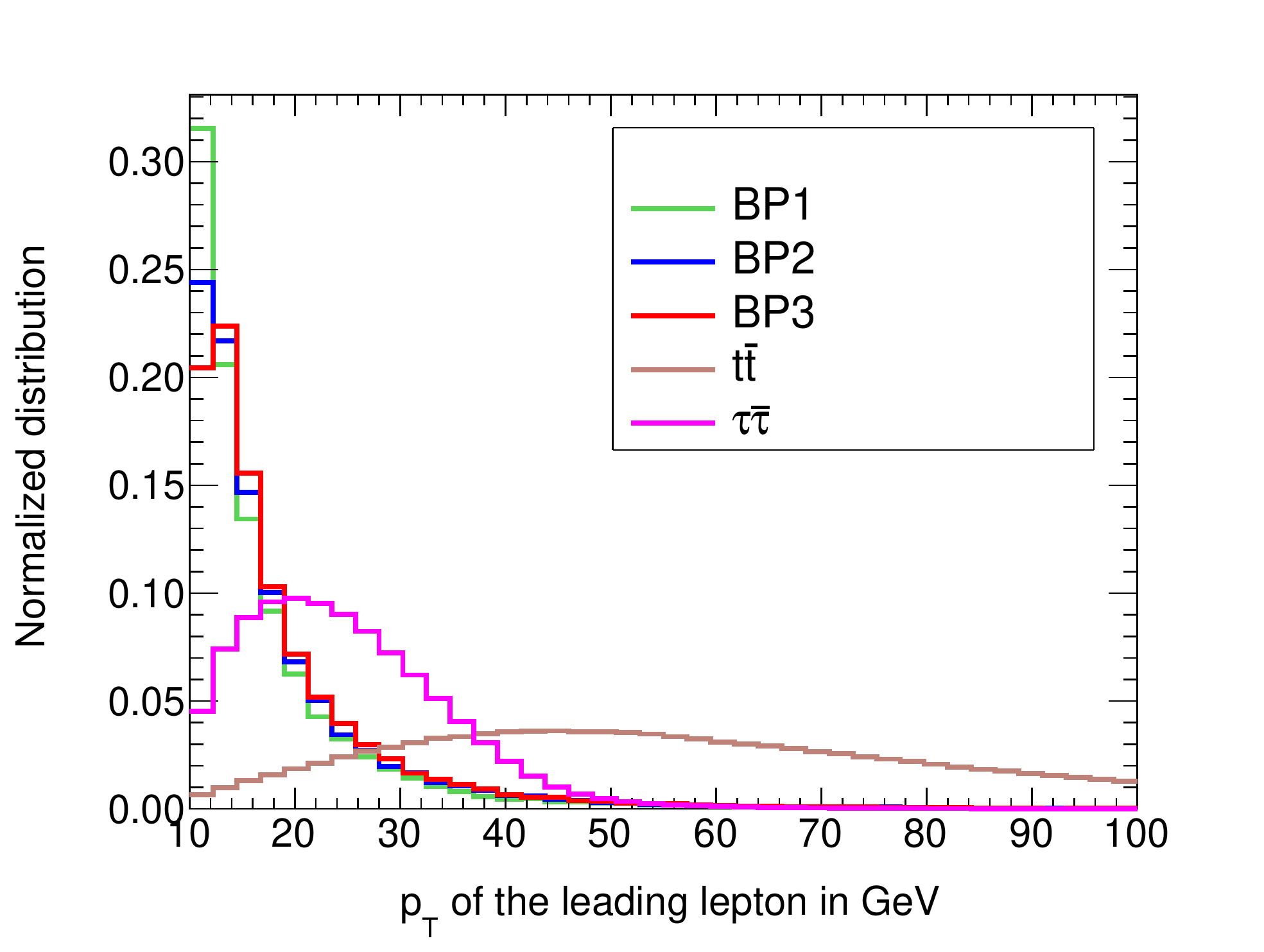}
	\includegraphics[width=7.2cm,height=6.0cm]{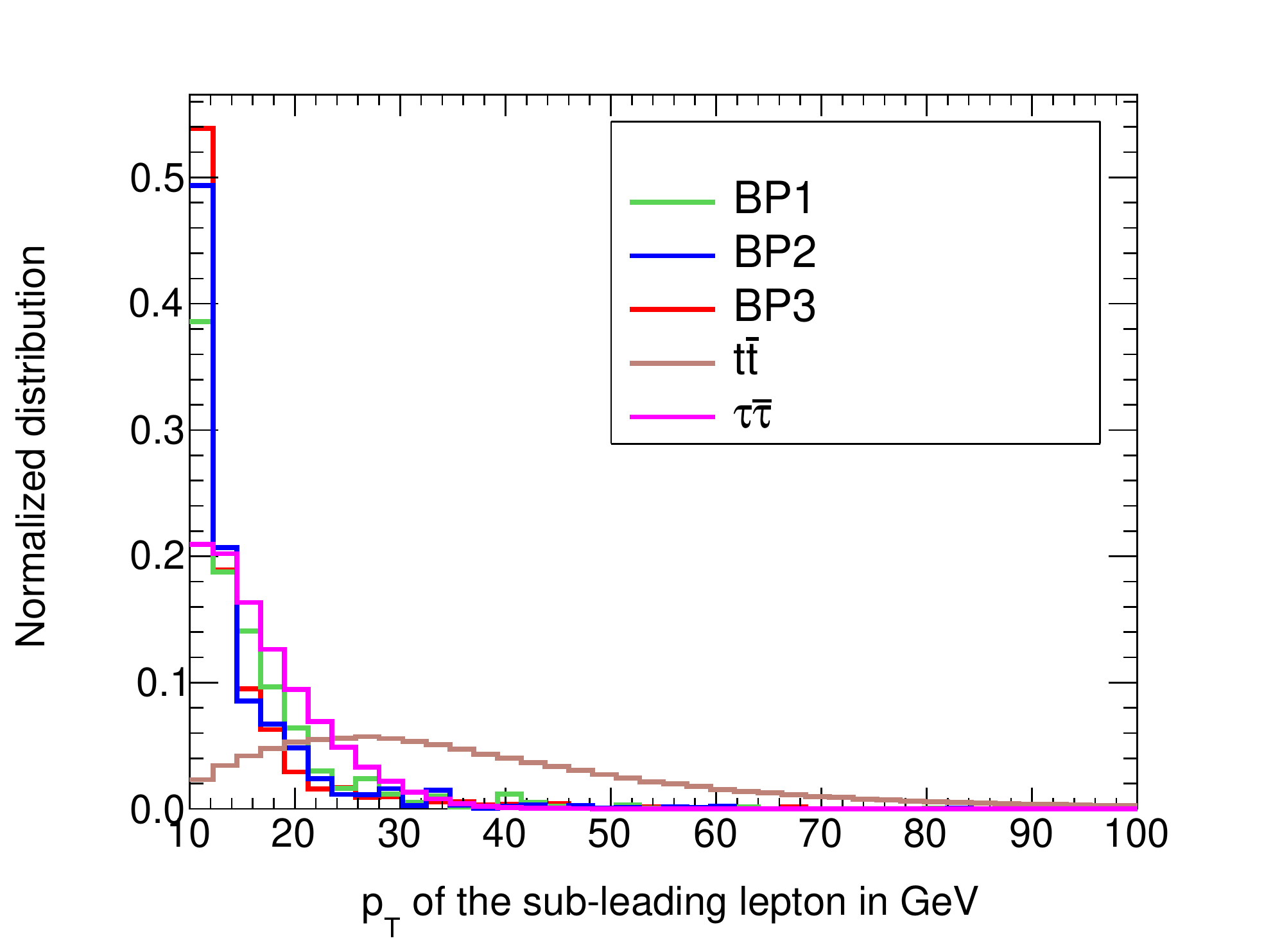}
\caption{\it Distribution of transverse momenta of leading (left) and sub-leading (right) leptons for signal and backgrounds.}
\label{ptlep}
\end{figure}

\begin{itemize}
           \item {\bf $p_T$ of the leptons:} In Fig.~\ref{ptlep}, we present the transverse momentum $p_T$ of the leading and sub-leading leptons. For the signal, the leptons coming from the decay of a low mass pseudoscalar, tend to have low $p_T$. Since the distributions are mostly overlapping for both signal and $\tau \tau$ background, it is very difficult to put any hard $p_T$ cut. However, to affirm that our signal has 2 isolated leptons, we reject any third lepton with $p_T(\ell) >$ 10 GeV. 
Moreover, since our signal is hadronically quiet, we put a jet-veto of with $p_T(j) >$ 20 GeV. We also reject any 
$b$ - jet with $p_T(b) >$ 20 GeV. This is our preselection cut as described in Table~\ref{tab:sig}. This helps us reduce the $t\bar{t}$ semileptonic and $W^{\pm}+$ jets background.

\begin{figure}[!hptb]
	\centering
	\includegraphics[width=7.2cm,height=6.0cm]{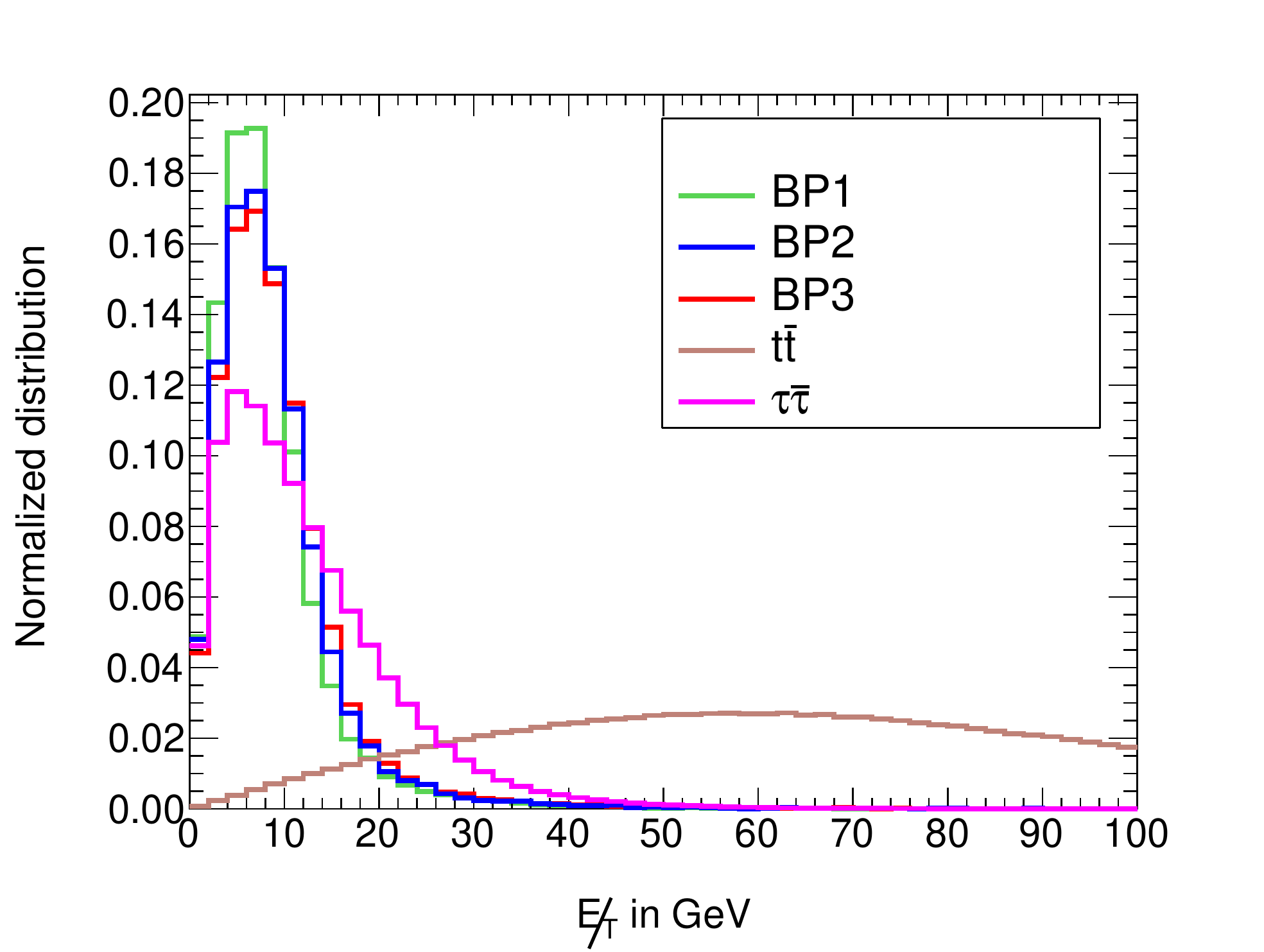}
	\includegraphics[width=7.2cm,height=6.0cm]{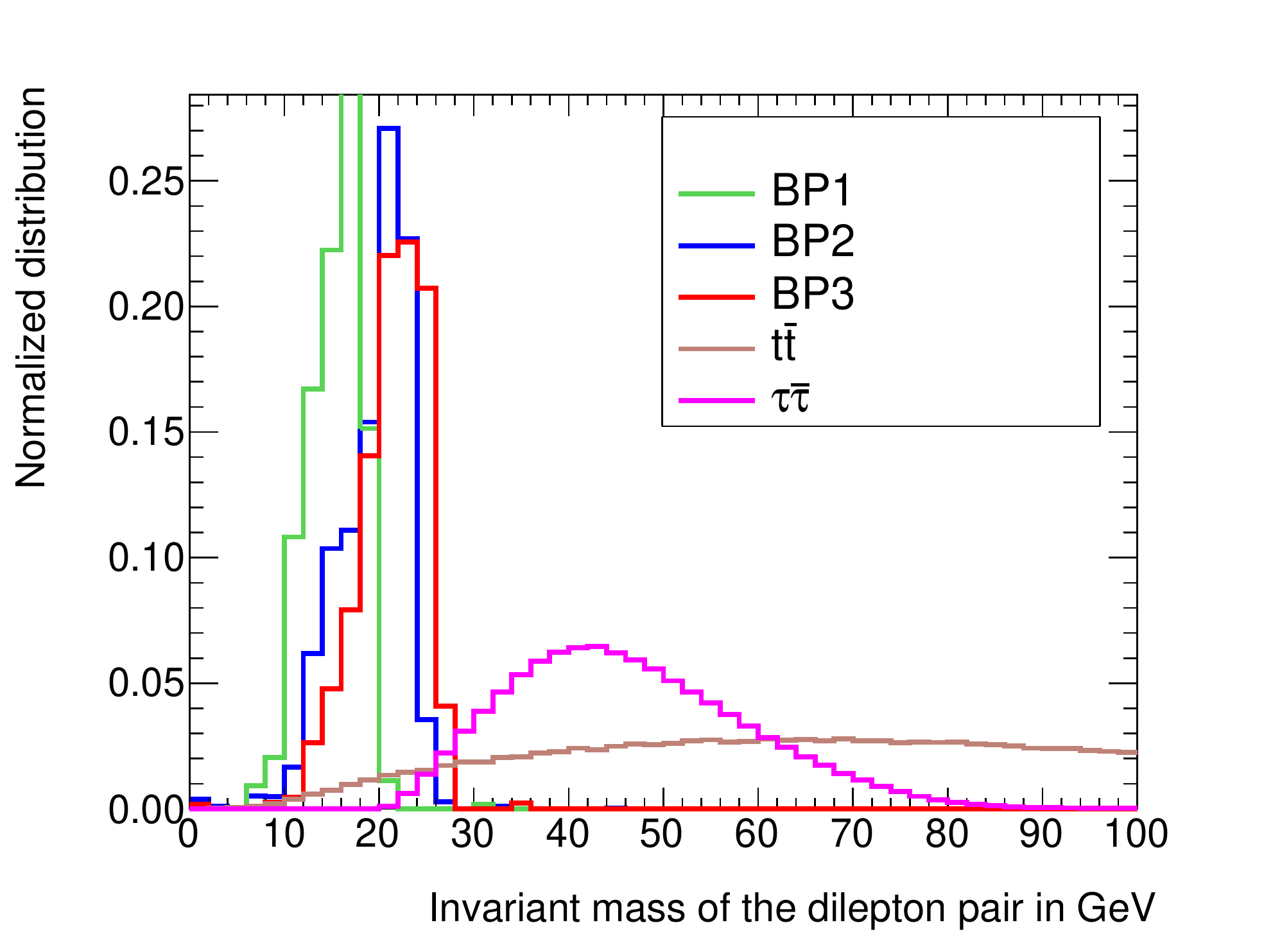}
\caption{\it Distribution of $\slashed{E_T}$ (left) and invariant mass of two leptons for signal and backgrounds.}
\label{met_invll}
\end{figure}

\begin{figure}[!hptb]
	\centering
	\includegraphics[width=7.2cm,height=6.0cm]{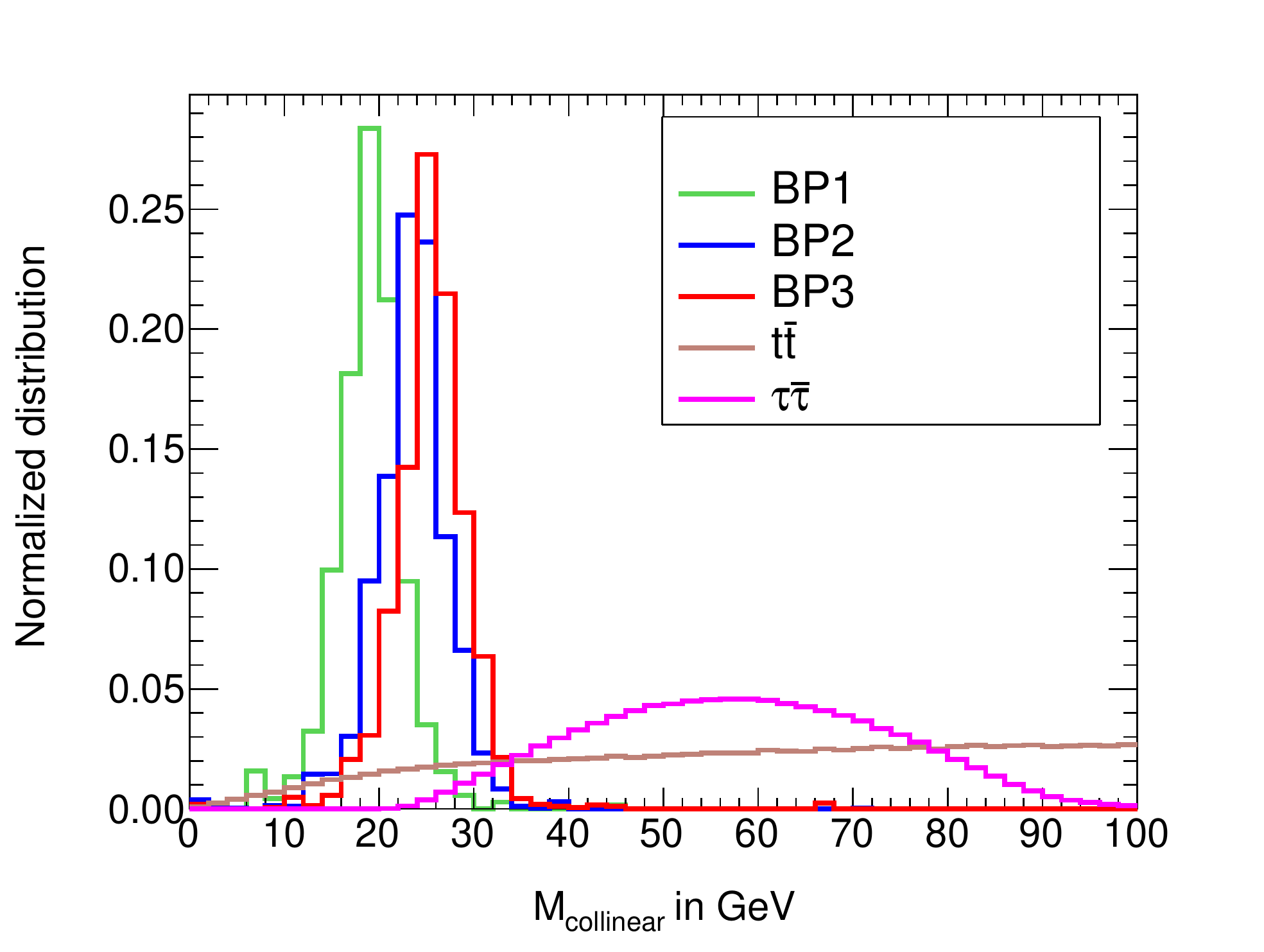}
	\includegraphics[width=7.2cm,height=6.0cm]{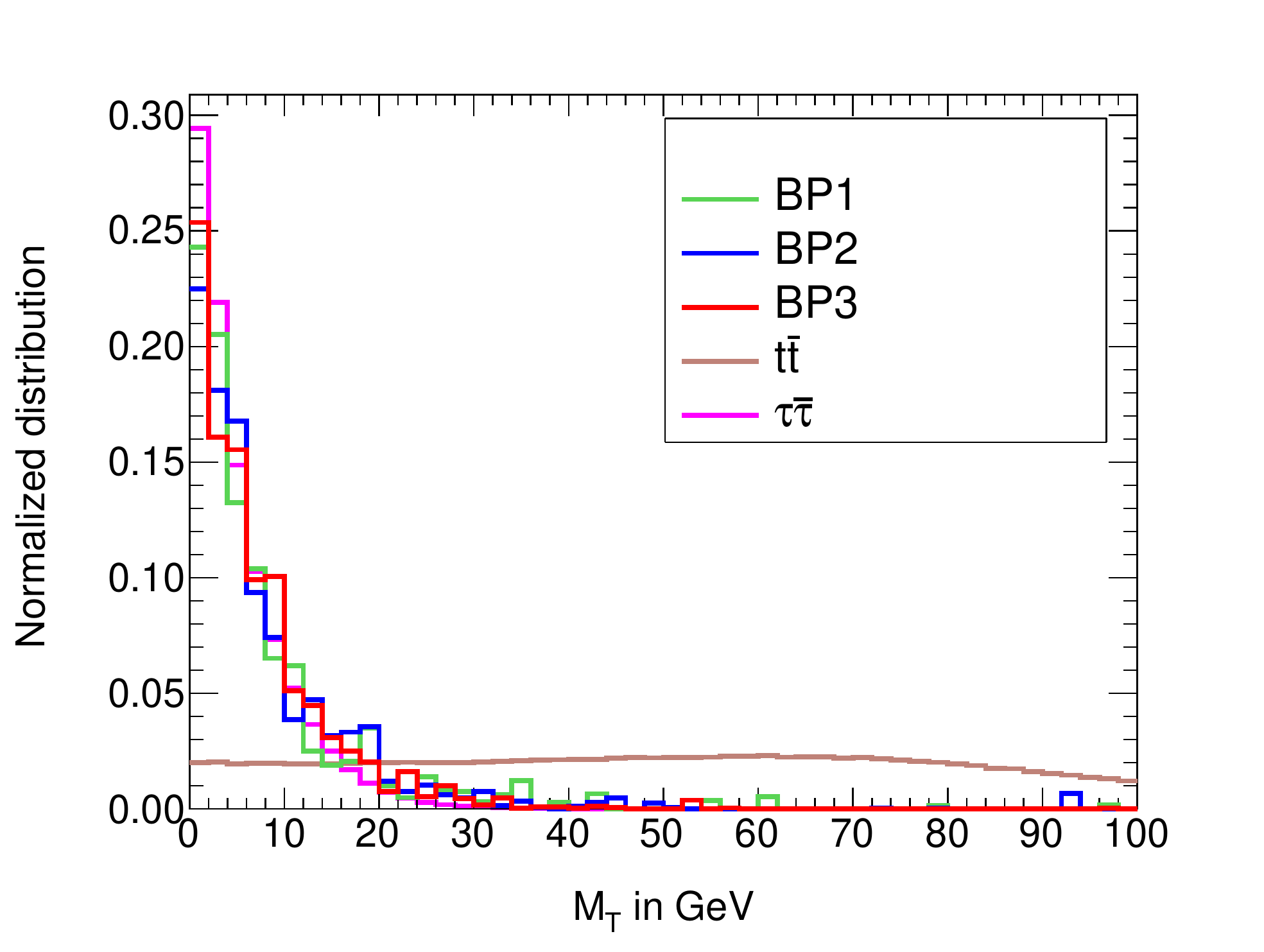}
\caption{\it Distribution of collinear mass (left) and transverse mass (right) for signal and backgrounds.}
\label{mcollinear_mtransverse}
\end{figure}

\begin{figure}[!hptb]
	\centering
	\includegraphics[width=7.2cm,height=6.0cm]{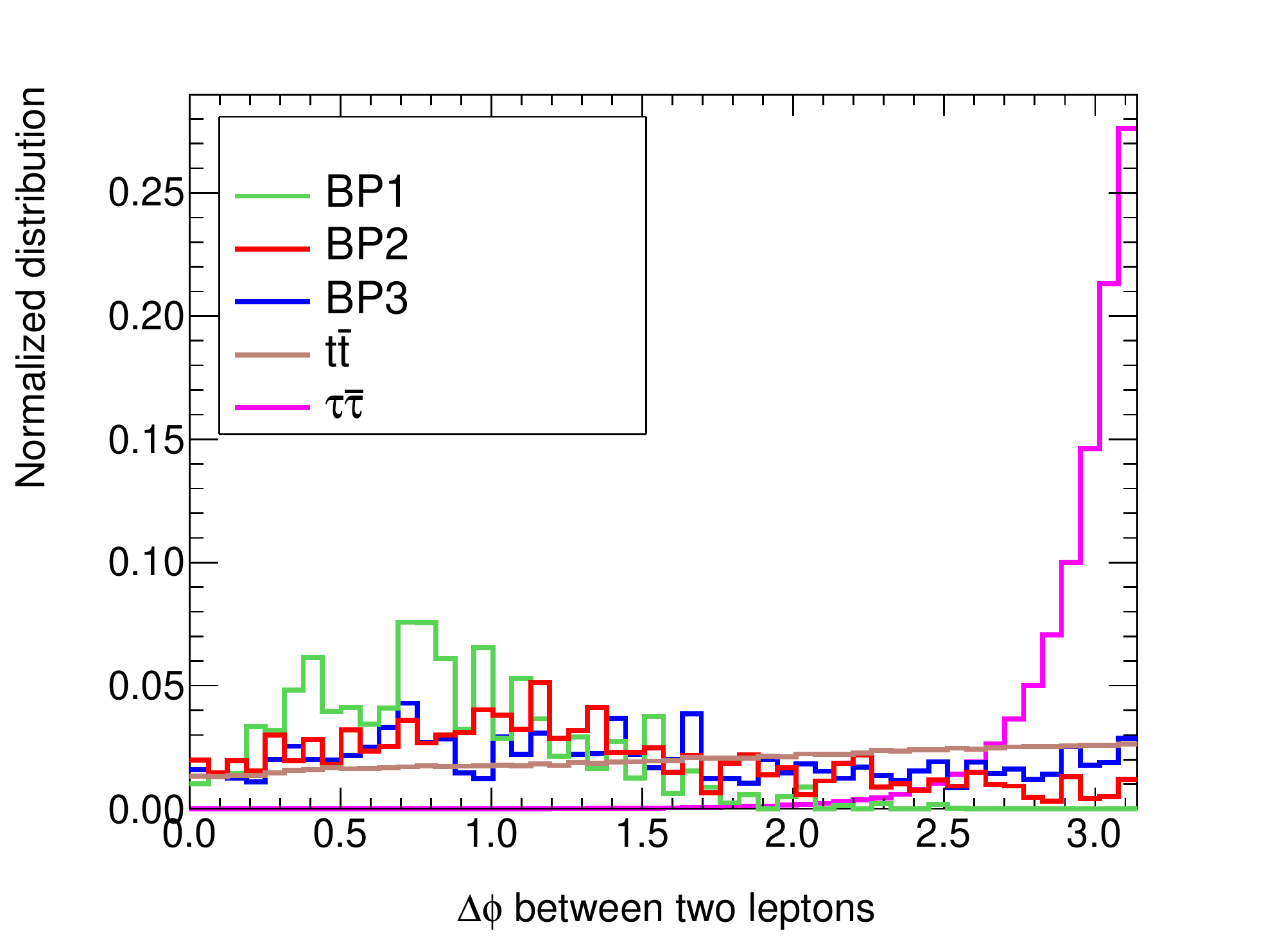}
\caption{\it Distribution of $\Delta \phi$ between two leptons for signal and backgrounds.}
\label{delpill}
\end{figure}

           \item {\bf Selecting $\slashed{E_T}$:} For signal the neutrino is coming from the leptonic decay of $\tau$. The $\tau$ comes from the decay of low mass pseudoscalar. Therefore for signal the $\slashed{E_T}$ tends to be small. For $\tau \tau$ background, the neutrinos are produced almost back to back. So, the lower $\slashed{E_T}$ bins are populated both for signal and $\tau \tau$ background. On the other hand, top decay being a three-body decay, the $\slashed{E_t}$ produced in $t \bar t$ event peaks at a larger value. We present the distribution of $\slashed{E_T}$ in Fig.~\ref{met_invll}(left).

   \item {\bf Invariant mass of the di-lepton pair:}  In Fig.~\ref{met_invll}(right) we show the invariant mass of the di-lepton pair $M_{\ell\ell'}$. In case of signal, the leptons are coming from the decay of a low mass pseudoscalar, therefore its distribution peaks at a lower value, compared to $\tau\tau$ and $t \bar t$ background.  We mention here that this observable plays a crucial role in reducing the $ee/\mu\mu$ background to a large extent. The invariant mass for $ee/\mu\mu$ peaks at a $Z$-boson mass and therefore a suitable cut on this variable helps us get rid of this background. In addition, $M_{\ell\ell'}$ turns out to be an important observable to discriminate between $\tau\tau$ background and signal.

 \item {\bf The collinear mass:}  The collinear mass is defined as follows:
       \begin{equation}
        m_A = M_{collinear} = \frac{M_{vis}}{\sqrt{x_{\tau_{vis}}}},
       \end{equation}
       with the visible momentum fraction of the $\tau$ decay products being,
       $x_{\tau_{vis}}=\frac{|\vec{p}_T^{\; \tau_{vis}}|}{|\vec{p}_T^{\; \tau_{vis}}|+|\vec{p}_T^{\; \nu}|}$, where
       $\vec{p}_T^{\; \nu}=|\vec{\slashed{E}}_T| \hat{p}_T^{\; \tau_{vis}}$ and $M_{vis}$ is the visible mass of the $\tau - \ell$ system. 
The variable $M_{collinear}$ essentially reconstructs the mass of the pseudoscalar. From Fig.~\ref{mcollinear_mtransverse} (left) it is clear that $M_{collinear}$ yields a clear distinction between the signal and the irreducible $\tau \tau$ background. A suitable cut should be imposed on this variable to reduce the $\tau\tau$ background.

     \item {\bf The transverse mass variable:}  The transverse mass is defined as
       \begin{equation}
          M_T(\ell) = \sqrt{2 p_T(\ell)\vec{\slashed{E}}_T (1 - \cos \Delta \phi_{\vec{\ell}-\vec{\slashed{E}}_T})}
       \end{equation}    
where $ \Delta \phi_{\vec{\ell}-\vec{\slashed{E}}_T}$ denotes the azimuthal angle between the leading lepton and $\slashed{E}_T$. From Fig.~\ref{mcollinear_mtransverse} (right) it is clear that a cut on $M_T$ variable helps us reduce the $t\bar{t}$ background.
           
   \item {\bf Angle between the lepton:}  The angle between two leptons is strictly correlated to the invariant mass. Since for the signal the invariant mass of the di-lepton pair is small, the azimuthal angle between the two leptons $\Delta\phi_{\ell\ell'}$ appears at lower value compared to the $\tau \tau$ background where the leptons are produced back to back and as a result $\Delta\phi_{\ell\ell'}$ peaks around $\pi$.
It is clear from Fig.~\ref{delpill} that a suitable cut on this observable will enhance the signal-background separation.

          \end{itemize}

 \begin{table}[ht!]
	\centering
	\scriptsize
	\begin{tabular}{|p{2.0cm}|c|c|c|c|c|c|}
		\cline{2-7}
		\multicolumn{1}{c|}{}& \multicolumn{6}{|c|}{Effective NLO cross-section after the cut(fb)}  \\ \cline{1-7}
		SM-background  
		& Preselection cuts &$\Delta\phi_{\ell\ell'} < 2.2$  & $M_{\ell\ell'}$ $ < 15$ GeV & $\slashed{E_T} < 15$ GeV  &  Mcollinear $> 10$ GeV & $M_T < 25$ GeV
		\\ \cline{1-7} 
                  $\tau\tau$ & 8582.75 & 132.089 & 0.21 & 0.089 & 0.052 & 0.052  \\ \cline{1-7} 
		$t\bar{t}$ leptonic & 22.10  & 11.01 & 0.099 & 0.016 & 0.016 & 0.0016  \\ \hline \hline
			\multicolumn{1}{|c|}{Signal }  &\multicolumn{6}{|c|}{}   \\ \hline
		\multicolumn{1}{|c|}{BP1}&  0.0689  & 0.0686 & 0.0276 & 0.0266 & 0.0262 & 0.0258  \\ \hline
		\multicolumn{1}{|c|}{BP2}& 0.0637 & 0.0542 & 0.0081  & 0.0076  & 0.0073 & 0.0073 \\ \hline 
                     \multicolumn{1}{|c|}{BP3}& 0.0513 & 0.0381 & 0.0028 & 0.0026 & 0.0025 & 0.0025 \\ \hline
		
	\end{tabular}

          \begin{tabular}{|c|c|}
         \hline 
      Benchmark points &  Significance reach at 3 $ab^{-1}$ luminosity \\ \hline
      \hline 
  BP1 & 5.7 $\sigma$ \\ \hline
  BP2 & 1.7 $\sigma$\\ \hline
  BP3 & 0.6 $\sigma$\\ \hline
\hline
 \end{tabular}
	\caption{\it The cut-flow for signal and background and significance reach for our signal at 14 TeV LHC for 3 $ab^{-1}$ luminosity. }
	\label{tab:sig}
\end{table}

With optimized cuts on the aforementioned variables (listed in Table~\ref{tab:sig}), we get the signal significance as presented in Table~\ref{tab:sig}. The significance~\cite{Cowan:2010js} is calculated using the following formula.
${\cal{S}}=\sqrt{2[(S+B) ln(1+ \frac{S}{B}) -S]}$

where $S$ and $B$ denotes the number of signal and background after applying all the cuts. We mention here that, we multiply the signal and background cross-sections with respective {\it{k-factors}} to take into account the 
next-to-leading-order (NLO) effects. For signal, we use the $k$-factor of 2~\cite{Djouadi:2003jg} while for $t\bar{t}$ and $\tau\tau$ background, we take the $k$-factor to be 1.6~\cite{Sirunyan:2017uhy} and 1.15~\cite{Catani:2009sm} respectively.


\subsection{Improved analysis with Artificial Neural Network (ANN)}

Having performed the cut-based analysis, we proceed to analyze the same channel(di-lepton + $\slashed{E_T}$) with ANN~\cite{Teodorescu:2008zzb} at the LHC. We explore the possibility of improvement in our results. This method has been used extensively in the recent past and it has proved to improve the results of cut-based analysis multifold enabling better separation between the signal and backgrounds. Significant work has been done in the context of Higgs sector~\cite{Hultqvist:1995ibm,Field:1996rw,Bakhet:2015uca,Dey:2019lyr,Lasocha:2020ctd}. In collider search for dark matter also this methods have been proved to be extremely useful~\cite{Dey:2019lyr,Dey:2020tfq}. Therefore we employ this tool in our present analysis also where signal yield is small because of minuscule branching fraction ($\sim 10^{-7}$) of the flavor changing decay of the pseudoscalar and the distinction between signal and background becomes crucial. We have examined and computed the maximum signal significance for the benchmarks that we considered, achievable at the HL-LHC using these technique. The toolkit used for ANN analysis is a python-based deep-learning library Keras~\cite{keras}.

From our analysis in the previous section we identify the important input variable that provide substantial distinction between signal and backgrounds. We mention here that the choice of input variables play a crucial role. In Table.~\ref{featurevar} we present all the input variables that we have used for our analysis.

\begin{table}[htpb!]
\centering

\begin{tabular}{||c | c||} 
 \hline
 Variable & Definition \\ [0.5ex] 
 \hline\hline
 $p^{\ell_1}_{T}$ & Transverse momentum of the leading lepton \\ 
 $p^{\ell_1}_{T}$ & Transverse momentum of the sub-leading lepton \\
 $E^{miss}_{T}$ & Missing transverse energy \\
 $M_{\ell\ell'}$ & Invariant mass of the di-lepton pair \\
 $\Delta \phi_{\ell\ell'}$ & Azimuthal angle difference between the di-lepton pair \\ 
 $\Delta R_{\ell\ell'} $& $\Delta R$ separation between the di-lepton pair \\
 $M_{vis}$ & Visible mass of the di-lepton system \\
 $x_{vis}$ & Visible momentum fraction of the $\tau$ decay products \\
 $M_{collinear}$ & Collinear mass \\
 $M_T$ & Transverse mass \\
 $\Delta \phi_{\ell_1 \slashed{E_T}}$ & Azimuthal angle difference between the leading lepton and $\slashed{E_T}$ \\
 $\Delta \phi_{\ell_2 \slashed{E_T}}$ & Azimuthal angle difference between the sub-leading lepton and $\slashed{E_T}$ \\[1ex] 
 \hline
 \end{tabular}

 \caption{\it Feature variables for training in the ANN analysis.}
  \label{featurevar}
\end{table}

For ANN analysis we have chosen a network with four hidden layers with activation curve relu at all of them. The batch-size is taken to be 1000 and the number of epochs is 100 in our case for each batch. We have used 80\% of the dataset for training and 20\% for validation. One  possible demerit of these techniques is over-training of the data sample. In case of over-training the training sample indeed gives extremely good accuracy but the validation sample fails to achieve the same degree of accuracy as that of the training sample. However we have explicitly checked that with our choice of network parameters the algorithm does not over-train. 

\begin{figure}[!hptb]
	\centering
	\includegraphics[width=9cm,height=7.0cm]{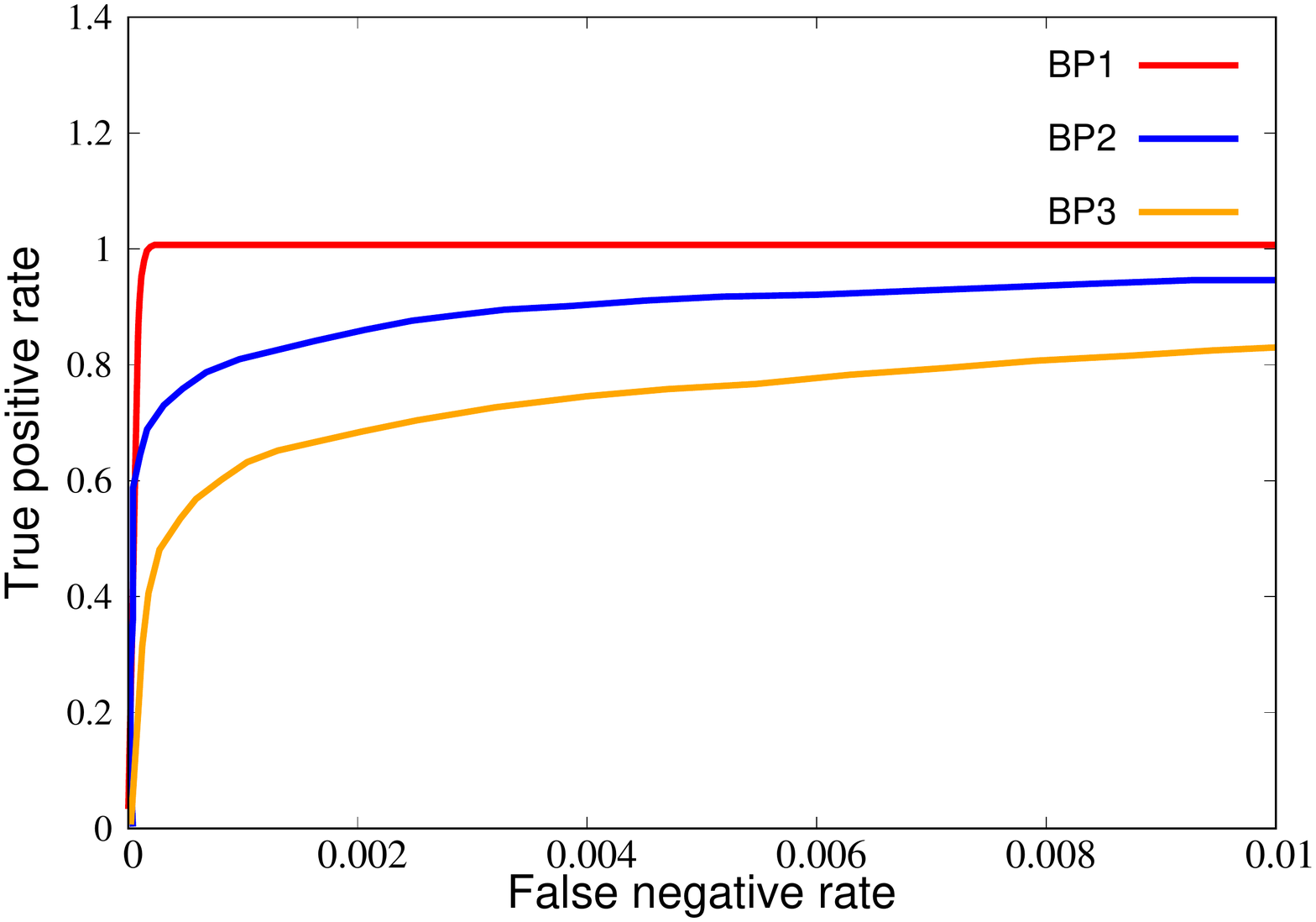}
\caption{\it ROC curves for BP1, BP2 and BP3}
\label{roc}
\end{figure}

We find that the variables $M_{\ell\ell'}$, $M_{collinear}$, $M_T$, $\Delta \phi_{\ell\ell'}$ and $\Delta R_{\ell\ell'} $ play crucial role in signal-background separation. However, there is a strong correlation between $\Delta R_{\ell\ell'} $, $\Delta \phi_{\ell\ell'}$ and $M_{\ell\ell'}$ as we have already discussed in the previous subsection. We mention here that in order to obtain a better performance from the network we have applied two basic cuts, namely $M_{\ell\ell'} < 30$ GeV and $M_{collinear} < 40$ GeV on signal and background events over and above the lepton selection and jet-veto. From our discussion of the cut-based analysis we know that these cuts guide us towards the so-called signal region. Therefore the network will be trained better to separate signal from background specifically in the signal region, this results in a better accuracy in the output. The accuracy we get is 99\%(BP1), 98\%(BP2) and 96\%(BP3) which indicates very good discriminating power between signal and background. We present in Fig.~\ref{roc}, the Receiver Operating Characteristic (ROC) curve for the three benchmark points of our choice.

The area under curve is 0.999(BP1), 0.998(BP2) and 0.994(BP3). We plot only a part of the ROC curve which is relevant for our analysis. We scan over the ROC curve and choose suitable points which yield the maximum signal significance. In Table.~\ref{significance_ann}, we present the signal significance ${\cal S}$ for all the signal benchmarks we have considered.

\begin{table}[!hptb]
\begin{center}
\begin{normalsize}
\begin{tabular}{| c | c | }
\hline
BP & $
{\cal S}$ (cuts+ANN)   \\
\hline
BP1  & 9.2 $\sigma$ \\
\hline
BP2  & 5.3 $\sigma$  \\
\hline
BP3  & 3.2 $\sigma$  \\
\hline
\end{tabular}
\end{normalsize}
\caption{\it Signal significance for the benchmark points at 14 TeV with ${\cal L}$ = 3 $ab^{-1}$ with cuts+ANN. }
\label{significance_ann}
\end{center}
\end{table}

Comparing the results of ANN in Table.~\ref{significance_ann} and the cut-based results in Table.~\ref{tab:sig} we can see that our analysis with ANN improves the results of cut-based analysis significantly.

\section{Conclusion}\label{conclusion}

The motivation behind this work is a much-desired reconciliation between the observed muon anomaly and LFV constraints. In this regard, we consider an extension of the SM with extended scalar sector, namely, generalized 2HDM. The additional non-standard 
scalars of this model take part in the muon anomaly and flavor non-diagonal Yukawa matrices lead to LFV processes. We show that the long-standing problem of muon anomaly and LFV constraints can be solved simultaneously over considerable range of parameter space in this model. We show such a region in Fig.~\ref{case4} with flavor changing couplings fixed at $y_{\mu e}= 10^{-7}, y_{\tau e}=5\times 10^{-5}$ and $y_{\mu \tau}=5\times 10^{-5}$ and the non-standard CP-even and charged Higgs masses are fixed at 120 GeV and 150 GeV respectively, where both muon anomaly and LFV constraints are satisfied.

We then proceed to implement theoretical constraints pertaining to the requirement of perturbativity, unitarity and vacuum stability. The flavor non-diagonal Yukawa matrices also get severely constrained by the $B$-physics observables. The direct searches for the SM Higgs as well as the additional scalar states in the collider put another set of bounds on the model parameter space. One such crucial direct search constraint turns out to be the search for the SM Higgs decaying to two light pseudoscalars. Our aim in this work is to search for lepton flavor violation in the scalar sector at the collider. Therefore the scalar states with low mass, prove to be the best candidate for such searches owing to their large production cross-section. We have also found that it is the light CP-odd scalar($A$) of our model that helps us explain the $(g_{\mu}-2)$ data. The lightness of the pseudoscalar, also implies a large branching ratio of the 125 GeV Higgs into a pair of pseudoscalars when the decay is kinematically feasible. To ensure the upper bound to this branching fraction coming from collider data, along with the perturbativity requirements, one is drawn to the scenario where the observed 125 GeV Higgs is the heavier of the two CP-even states of 2HDM in the alignment limit. However, this statement is valid only in the `right-sign' region of 2HDM which we have considered in this work. The `wrong-sign' scenario will give rise to a different allowed region and interesting phenomenology, which we want to pursue in detail in the future.

After finding out the region allowed by all constraints, we look for flavor violating decay of CP-odd scalar $(A$) in the $ \ell \tau \rightarrow \ell^+ \ell'^- + \slashed{E_T}$ final state, where $\tau$ decays leptonically and $\ell, \ell'= e, \mu$. 
Performing a rectangular cut-based analysis for $14~\rm{TeV}$ LHC with $3 ab^{-1}$ luminosity,  we show that the significance drops from $\sim 6  \sigma ~\rm{to} \sim 1\sigma$ as the mass of the scalar increases from $21~\rm{ GeV }~\rm{to}~ 27 ~\rm{GeV}$. 
We then perform an analysis using ANN and observe significant improvements of our results from the cut-based analysis. We would like to point out that although we have probed a narrow region of parameter space in terms of pseudoscalar mass, we did it in order to investigate the reach of LFV searches at 14 TeV LHC with 3 $ab^{-1}$ at $\gtrsim 3\sigma$ significance. The results of ANN analysis in Table.~\ref{significance_ann} indicates that even higher masses of $A$ can be probed, although with somewhat lower significance ($< 3\sigma$). Also, a further upgrade in the luminosity as well as energy frontier will enable us to probe heavier CP-odd scalar decaying into lepton flavor violating final states.


\section{Acknowledgement}

This work was supported by funding available from the Department of Atomic Energy,  Government of India, for the Regional Centre for Accelerator-based Particle Physics (RECAPP), Harish-Chandra Research Institute.



\providecommand{\href}[2]{#2}\begingroup\raggedright\endgroup

\end{document}